\documentclass[%
 reprint,
 amsmath,amssymb,
 aps,
 prl,
 superscriptaddress
]{revtex4-2}

\usepackage{graphicx}
\usepackage{dcolumn}
\usepackage{bm}
\usepackage{textcomp, gensymb}
\usepackage{amsmath}
\usepackage{mathbbol}
\usepackage{amssymb} 
\begin{document}

\preprint{APS/123-QED}

\title{Coherent Perfect Absorption of Arbitrary Wavefronts\\ at an Exceptional Point}

\author{Helmut Hörner}
\email{helmut.hoerner@tuwien.ac.at}
\affiliation{Institute for Theoretical Physics, TU Wien, Wiedner Hauptstra\ss e 8-10/136, A-1040 Vienna, Austria}

\author{Lena Wild}
\affiliation{Institute for Theoretical Physics, TU Wien, Wiedner Hauptstra\ss e 8-10/136, A-1040 Vienna, Austria}

\author{Yevgeny Slobodkin}
\affiliation{Institute of Applied Physics, The Hebrew University of Jerusalem, Bergman 206, 9190401, Jerusalem, Israel}

\author{Gil Weinberg}
\affiliation{Institute of Applied Physics, The Hebrew University of Jerusalem, Bergman 206, 9190401, Jerusalem, Israel}

\author{Ori Katz}
\email{orik@mail.huji.ac.il}
\affiliation{Institute of Applied Physics, The Hebrew University of Jerusalem, Bergman 206, 9190401, Jerusalem, Israel}

\author{Stefan Rotter} 
\email{stefan.rotter@tuwien.ac.at}
\affiliation{Institute for Theoretical Physics, TU Wien, Wiedner Hauptstra\ss e 8-10/136, A-1040 Vienna, Austria}

\date{\today}

\begin{abstract}
A Coherent Perfect Absorber (CPA) exploits the interferometric nature of light to deposit all of a light field's incident energy  into an otherwise weakly absorbing sample. The downside of this concept is that the necessary destructive interference in CPAs gets easily destroyed both by spectrally or spatially detuning the incoming light field. Each of these two limitations has recently been overcome by insights from exceptional-point physics and by using a degenerate cavity, respectively. Here, we show how these two concepts can be combined into a new type of cavity design, which allows broadband exceptional-point absorption of arbitrary wavefronts. We present two possible implementations of such a Massively Degenerate Exceptional-Point absorber and compare analytical results with numerical simulations.
\end{abstract}

\maketitle


\begin{figure*}
\includegraphics[height=7cm, width=\textwidth]{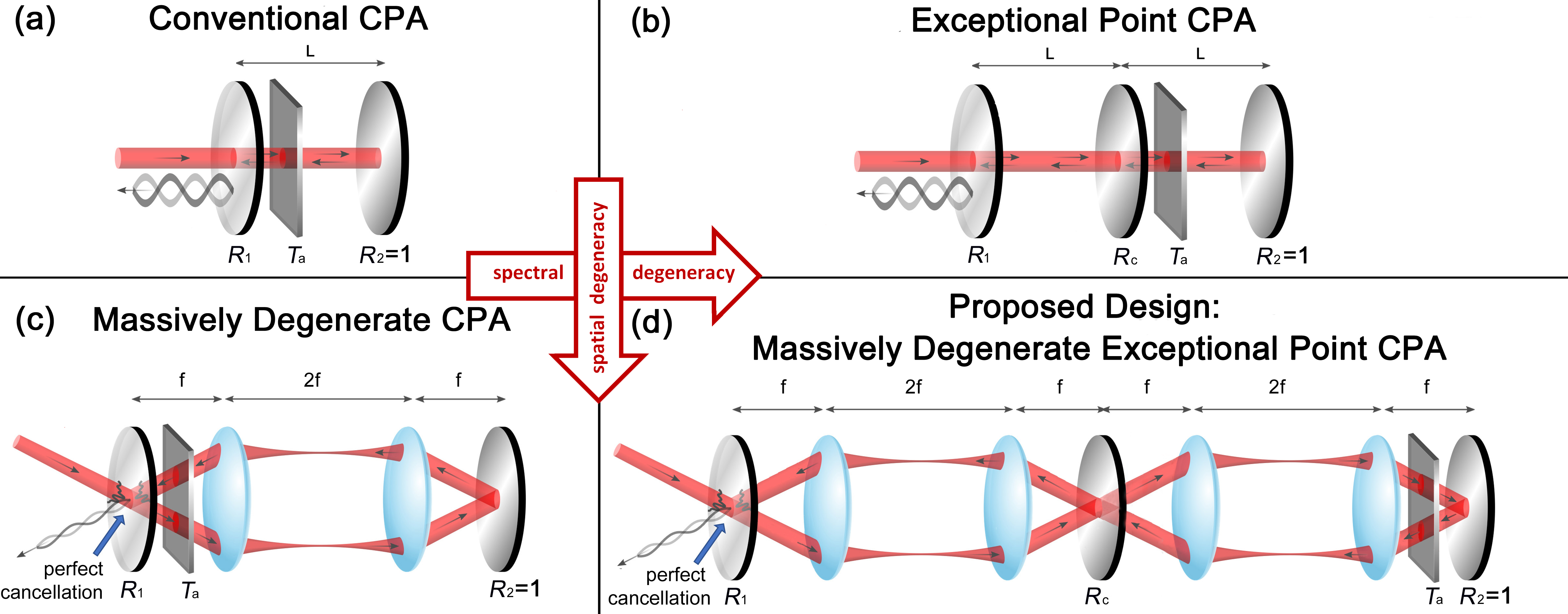}
\caption{\label{fig:CPA-comparisons} (a)~A conventional CPA (critically coupled cavity). When the critical-coupling condition $R_1=T_{\mathrm{a}}^2$ is met, the incident coherent light-field is perfectly absorbed, but only for a single plane-wave input mode at the right frequency. (b)~A spectrally degenerate EP-CPA, consisting of two weakly coupled conventional CPAs. When the critical-coupling-conditions $R_{\mathrm{c}}=4R_1/\left(1+R_1\right)^2$ and $R_1=T_{\mathrm{a}}^2$ are met, the (generally distinct) resonance-points merge at a single, real frequency. This  spectral degeneracy results in a broadened (quartic) absorption spectrum (not shown). (c)~A  MAD-CPA is a spatially-degenerate extension of the conventional CPA: Because of the self-imaging telescopic lens-arrangement, any incoming spatial mode is mapped onto itself after each round-trip, leading to perfect cancellation of any back-reflected light at the front mirror. 
(d)~The MAD-EP-CPA (``Design A'') combines the features of spatial and spectral degeneracies for improved spatial and spectral acceptance.}
\end{figure*}
{\it Introduction.}\,\textemdash\,While thin films or layers can only absorb a certain percentage of the light that shines through them in a single pass, an enhancement of absorption can be reached when the light beam has multiple passes through the absorbing material. This insight already indicates that placing an absorber in a cavity, where light bounces back and forth, can enhance the degree of absorption considerably (see Fig.~\ref{fig:CPA-comparisons}(a) as an example). The interferometric nature of light propagation in a cavity can boost the absorption even further~-- up to the point, where {\it all} the incoming light gets perfectly absorbed. In the single-mode case this phenomenon is known as ``critical coupling'' \cite{adler1969,Haus1984,Gorodetsky1999,Cai2000,yariv_2002}. For multiple modes one speaks of ``coherent perfect absorption", an effect that can also be interferometrically controlled through the relative phases of the incoming modes \cite{chong_coherent_2010, chong_hidden_2011, baranov_coherent_2017, Kishino1991, Uenlue1992, sweeney_perfectly_2019, Wang2021, Soleymani2022}. CPAs have been widely explored in diverse platforms and applications, 
encompassing complex structures and disordered media \cite{Jiang2023,Sweeney2020,Dhia2018,Horodynski2022,pichler_random_2019}, 
single-port interferometers \cite{Li2014,Jin2020}, 
optical switches \cite{Mock2012,Guo2023}, 
sensors \cite{Li2019,Zhang2022}, 
all-optical transistors and logic gates \cite{EbrahimiMeymand2020,Fang2015}. Recently, two of the major challenges of coherent perfect absorbers (CPAs) have been successfully addressed:

(i)~The first challenge is a CPA's narrow bandwidth, with slight frequency shifts disrupting the resonant absorption, typically resulting in a Lorentzian absorption profile. Research on how to make CPAs more broadband encompasses a wide array of approaches,
such as metasurfaces \cite{Guo2019,Zhang2022a,Zhang2022b},
plasmonic effects \cite{Pu2012,Baldacci2015,Wang2022},
metallic nanoparticles metasurfaces \cite{Noh2013},
thin films \cite{Li2015}, and broadband PCN resonators \cite{Choi2024}. 
Recently also insights from Exceptional Point (EP) physics have been used for advancing this goal \cite{sweeney_perfectly_2019, Wang2021,Soleymani2022,Zhong2020}: 
At an EP, two (or more) discrete eigenvalues and eigenstates of a non-Hermitian system merge. Such an EP-CPA can be observed when two (or more) cavities with spectrally-overlapping resonances are critically coupled as depicted in Fig.~\ref{fig:CPA-comparisons}(b). Specifically, it is possible to select system parameters such that both the critical-coupling condition, neccessary for perfect absorption, and the EP-condition coincide at the same real input frequency \cite{sweeney_perfectly_2019}. The result is a spectral degeneracy that leads to a significantly broadened (quartic) absorption spectrum \cite{Wang2021,Soleymani2022}, in analogy to the physics of white-light cavities \cite{Wicht1997, Yum2012, Pati2007, Smith2009, Smith2008, Kotlicki2014} and of multi-mirror Fabry-Pérot interferometers \cite{Stadt1985}.
 
 (ii)~The second major challenge concerns the limitation that a CPA typically only works for a well-defined input wavefront or mode. Detuning this input mode, e.g., in its incoming angle or phase profile, destroys the delicate interference necessary to achieve perfect absorption. To overcome this restriction, a massively degenerate (MAD) cavity turns out to be ideally suited \cite{Slobodkin2022}, as illustrated in Fig.~\ref{fig:CPA-comparisons}(c). In such a cavity, every incident light-field is always imaged onto itself after each round-trip. This configuration thus enables the necessary interferometric CPA-effect for any given spatial mode.   

The question we explore here is whether these two distinct concepts can be integrated into a single CPA design that simultaneously achieves \textit{spectral} and \textit{spatial} degeneracy. Our approach to this new cavity design is based on the following considerations:  A conventional CPA, as depicted in Fig.~\ref{fig:CPA-comparisons}(a), can be transformed into a spectrally degenerate CPA by critically coupling it to another cavity, as illustrated in Fig.~\ref{fig:CPA-comparisons}(b). Similarly, the same conventional CPA can be transformed into a spatially degenerate CPA by incorporating two lenses (focal length~$f$) in a telescopic configuration (of length~$4f$), as illustrated in Fig.~\ref{fig:CPA-comparisons}(c). Thus, merging these two ideas by critically coupling two spatially degenerate cavities, as illustrated in Fig.~\ref{fig:CPA-comparisons}(d), should result in a CPA that is both spatially and spectrally degenerate.

{\it Spatial and Spectral Degeneracy.}\,\textemdash\,To derive the reflection behavior of light incident from the left on this MAD-EP-CPA we employ a scalar optics model for polarized light, considering a large, yet finite, number of spatial modes. Let $\textbf{R}_1$, $\textbf{R}_2$, and $\textbf{R}_\mathrm{c}$ represent the reflection matrices, and $\textbf{T}_1$, $\textbf{T}_2$, and $\textbf{T}_\mathrm{c}$ the transmission matrices for the left, right and central mirror, respectively. Additionally,  $\textbf{T}_\mathrm{a}$ is the transmission through the absorber and $\textbf{T}_\text{4f}$ represents the transmission matrix for propagation through a $4f$-system, comprising two lenses in a telescopic configuration (without mirrors or absorbers). $\textbf{T}_\text{4f}$ acts as a double Fourier transform producing a spatially flipped image and a phase shift $e^{i \phi}$ that is the same for all modes. Using that the same reflection and transmission coefficients apply to all modes at the mirrors and the absorber, all matrices except $ \textbf{T}_\text{4f}$ become diagonal matrices, e.g.,  $\textbf{R}_1 = r_1 \mathbb{1}$, $\textbf{T}_1 = t_1 \mathbb{1}$, etc. Using a transfer matrix approach, we can construct from these individual transmission matrices the sought-after reflection matrix $\textbf{R}_{\text{\tiny{CPA}}}^{\text{\tiny{(A)}}}$ for the entire MAD-EP-CPA \cite[S1]{SM} (Design~A in Fig.~\ref{fig:CPA-comparisons}(d)):
\begin{align}
    \textbf{R}_\text{CPA}^\text{(A)} = \frac{t_\text{a} ^2 \, \textbf{T}_\text{4f}^4 +r_\text{c}   (1+r_1 t_\text{a}^2)  \,\textbf{T}_\text{4f}^2   + r_1 \mathbb{1}}{r_1 t_\text{a}^2 \, \textbf{T}_\text{4f}^4 + r_\text{c}(r_1+t_\text{a}^2)  \, \textbf{T}_\text{4f}^2 + \mathbb{1}} \label{eq:RCPA_simplified}
\end{align}
The matrix $\textbf{T}_\text{4f}^2$ in Eq.~(\ref{eq:RCPA_simplified}) results in a quadruple Fourier transformation with a uniform phase shift $e^{2i\phi}$ for all modes and can thus be represented as a simple diagonal matrix $\textbf{T}_\text{4f}^2=e^{2i\phi} \mathbb{1}$. Analogously, $\textbf{T}_\text{4f}^4=e^{4i\phi} \mathbb{1}$, turning $\textbf{R}_{\text{\tiny{CPA}}}^{\text{\tiny{(A)}}}$ into a diagonal matrix with identical entries $r_\text{CPA}(\omega)$ along the diagonal. Therefore, Eq.~(\ref{eq:RCPA_simplified}) can be reduced to a simple scalar equation, which proves that all modes are subject to the same reflection behavior in the system (\textit{spatial} degeneracy).
The function $r_\text{CPA}(\omega)$ is determined by the power-reflectivities $R_1=\vert r_1 \vert^2$, $R_\text{c} = \vert r_\text{c} \vert^2$ of the mirrors, and by the absorber's one-way transmissivity  $T_\text{a} = \vert t_\text{a} \vert^2$. To determine the conditions for \textit{spectral} degeneracy, i.e.~for the EP to occur at a real frequency, the values $R_1$, $R_\mathrm{c}$ and $T_\mathrm{a}$ need to be adjusted such that the two solutions $\omega_1$, $\omega_2$ of the equation $r_\mathrm{CPA}(\omega)=0$ merge into a single real-valued frequency $\omega$. 
These conditions are:
\begin{align}
R_1 =T_\mathrm{a}^2  \;\;\;\text{and}\;\;\;
R_\mathrm{c} = 
\frac{4 R_1}{\left(1 + R_1 \right)^2}
\label{eq:EP_conditions} 
\end{align}
A detailed derivation is available in the supplementary material \cite[S1-S3]{SM}.
The result (\ref{eq:EP_conditions}) shows that the relationship between $T_\mathrm{a}$ and $R_1$ for a MAD-EP-CPA is identical to that of a single-cavity MAD-CPA \cite{Slobodkin2022}: The higher the reflectivity $R_1$ of the input-coupling mirror, the weaker the internal absorber needs to be. Additionally, the relation between $R_\text{c}$ and $R_1$ indictates that $R_\text{c}$ quickly approaches values near $1$ with increasing values of $R_1$,  indicating weak coupling between the left and right sub-cavity. 
 
\begin{figure}
\includegraphics[height=7.5cm, width=8.6cm]{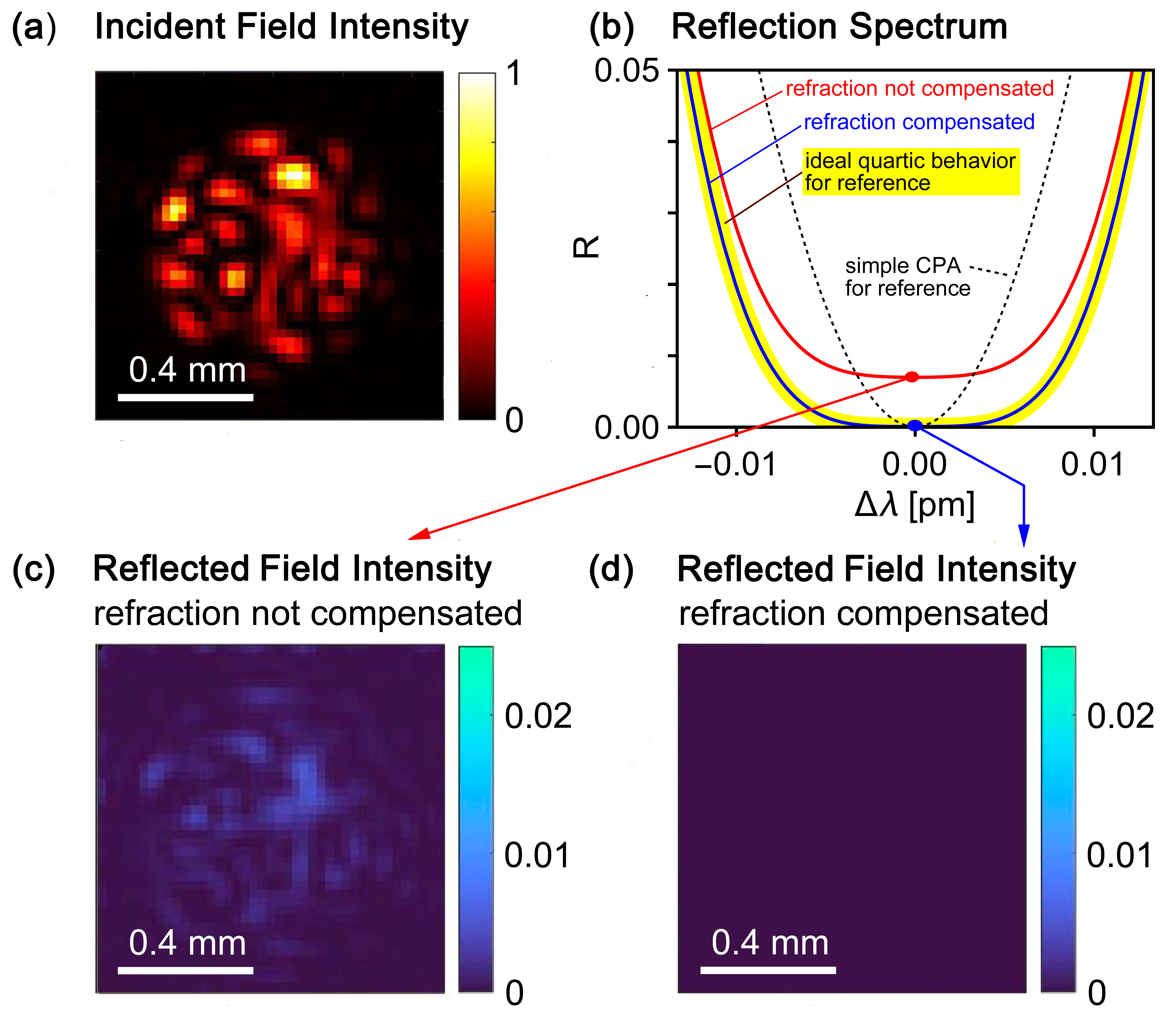}
\caption{\label{fig:num_sim} Numerical simulation of the proposed EP-MAD-CPA, as depicted in Fig.~\ref{fig:CPA-comparisons}(d). (a)~Intensity pattern of a simulated complex input random speckle field, composed of $160$ modes. (b)~Reflection spectrum across the entire field-of-view without compensation of the refraction aberrations caused by the absorber with thickness $0.6 \mathrm{mm}$ (red line), and with compensation of this shift included (blue line). For reference, ideal quartic behavior is shown with a thick yellow line, and a conventional CPA with the same absorber with a black dotted line. (c)~Reflected field intensity pattern at the resonance point when the refraction in the absorber is not compensated.  (d)~Reflected field intensity pattern at the resonance point with refraction compensated. (Simulation parameters: $R_1=0.7$, $R_\mathrm{c}=0.96886$.  $T_\mathrm{a}=\sqrt{0.7}$, wavelength $\lambda_0=633 \mathrm{nm}$, focal length $f=75\mathrm{mm}$.)
} 
\end{figure}

{\it Numerical Simulations.}\,\textemdash\,To corroborate that these analytical results are, indeed, applicable to the multi-mode MAD-EP-CPA, and to assess the impact of deviations in certain  system parameters (such as minor misalignments of the mirrors), we conducted  a corresponding numerical Scalar Fourier Optics computer simulation. As input we used a random speckle field, see Fig.~\ref{fig:num_sim}(a), generated by 160 modes of equal amplitude and random phase in $k$-space. Without compensating for the refractive index of the absorber (which we assumed to be $n_r=1.5$, at a thickness of $d=0.6\mathrm{mm}$), our numerical simulation already indicates a considerably broadened absorption lineshape, but with the minimum reaching only 0.7\%, see the red line in Fig.~\ref{fig:num_sim}(b). To address this, we adjusted the focal length of the rightmost lens in Fig.~\ref{fig:CPA-comparisons}(d) to $f'=f-d(n_\text{r}-1/n_\text{r})/2$. This adjustment compensates for the increased optical path-length caused by the absorber and preserves identical optical path-lengths in both the left and  right sub-cavity. As shown in the numerically calculated reflection spectrum of Fig.~\ref{fig:num_sim}(b) (blue line), this correction restores perfect absorption at the point of degeneracy and verifies the successful operation of our MAD-EP-CPA cavity design (see the broad yellow line in Fig.~\ref{fig:num_sim}(b) for comparison with an ideal quartic absorption lineshape). For a more detailed explanation of this refraction compensation technique, readers are referred to the supplementary text \cite[S8]{SM}. Another potential strategy to minimize refractive aberrations is to use an absorber with a refractive index whose real part is as close to $1$ as possible \cite{Tao2018}. 

\begin{figure}
\includegraphics[height=3.9cm, width=8.6cm]{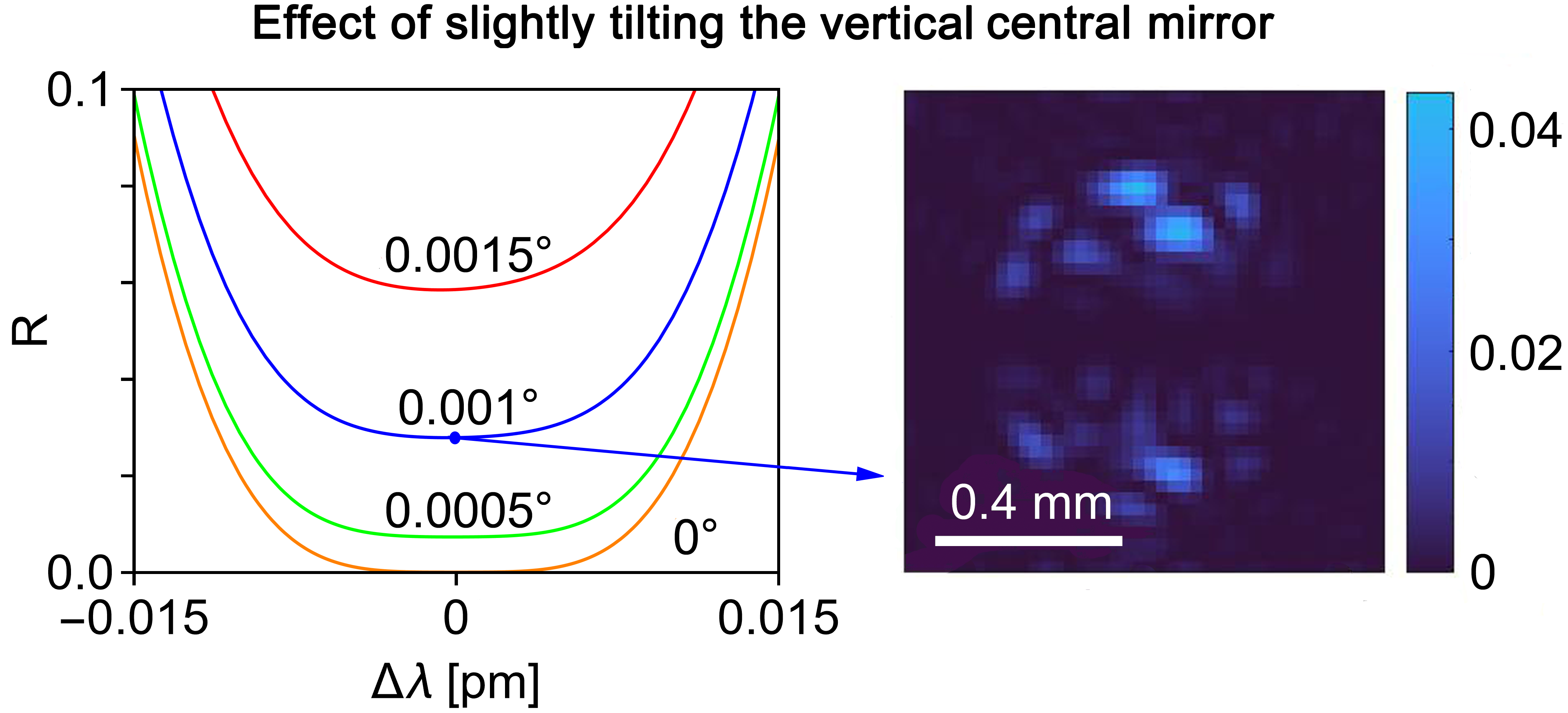}
\caption{\label{fig:num_sim2} Sensitivity to misalignment of the central coupling mirror (same simulation parameters as in Fig.~\ref{fig:num_sim}). \textit{Left:} Reflection spectrum across the entire field-of-view when the center mirror is slightly tilted out of the vertical alignment. \textit{Right:} Reflected field intensity pattern when the central mirror is tilted by $0.001^{\circ}$. Perfect absorption now only occurs in a narrow horizontal area along the tilt axis.} 
\end{figure}

The numerical simulations also allow us to study the impact of small deviations from optimal system parameters, and  help to pinpoint  the critical system parameters influencing the system performance. For instance, the simulations demonstrate  that the system exhibits robustness to deviations of the optimal reflectivity $R_1$ of the input coupling mirror (to within about $\pm 5\%$), or of the optimal transmissivity $T_\mathrm{a}$ of the absorber (to within about $\pm 2\%$).  In contrast, deviations in the reflectivity $R_\mathrm{c}$ of the central coupling mirror from its optimal value significantly affect the system performance. Even a mere $1\%$ deviation from the optimal value results in a minimum reflectivity of approximately $5\%$, significantly diminishing the system's absorption efficiency  (further details and figures can be found in the supplementary text \cite[S4]{SM}). As expected, the precise parallel alignment of the mirrors is also crucial, particularly for the central coupling mirror, as shown in Fig.~\ref{fig:num_sim2}. Additional details on critical parameters, such as the consequences of slight discrepancies in the alignment of the two cavity lengths or of residual lens reflections, are available in the supplementary text \cite[S4]{SM}. 

We have also studied MAD-EP-CPA systems encompassing more than two coupled cavities, as detailed in the supplementary text \cite[S5]{SM}.  Interestingly, constructing an EP-CPA with an increasing number of cavities results in a reflection spectrum around the resonance frequencies akin to a Butterworth bandpass filter, a concept from electrical filter theory \cite{Aatre2014, Butterworth1930}. This observation is consistent with the established understanding that, as outlined in \cite{Stadt1985}, the Butterworth function also characterizes the transmission behavior of a multi-mirror Fabry-Pérot interferometer in the absence of a dissipative element. 

{\it Alternative implementation.}\,\textemdash\,The initial design for a MAD-EP-CPA (``Design A''), as discussed thus far and illustrated in Fig.~\ref{fig:CPA-comparisons}(d), is probably the most obvious solution for merging the concepts shown in Fig.~\ref{fig:CPA-comparisons}(b)~and~(c). 
\begin{figure}
\includegraphics[height=7cm, width=8.6cm]{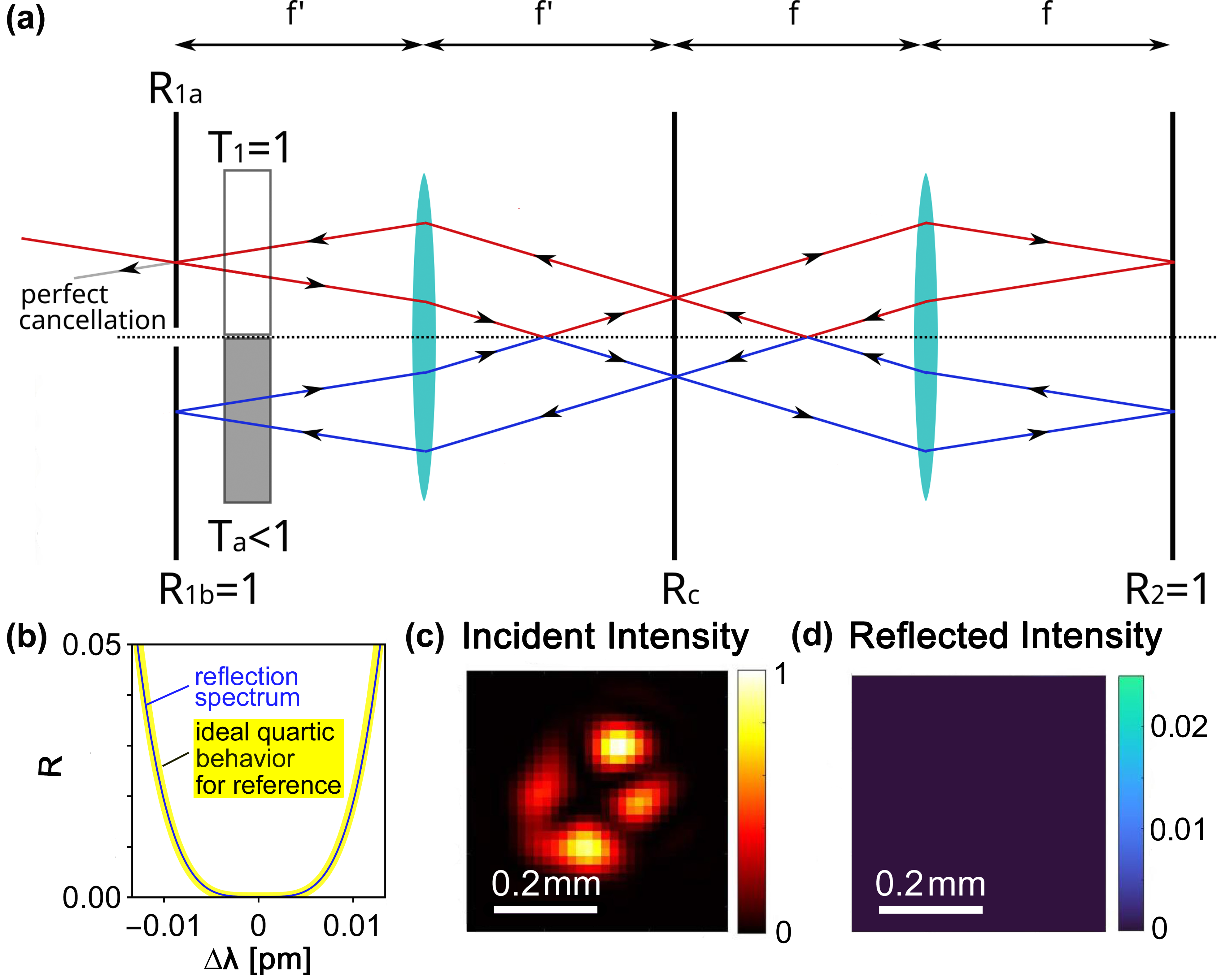}
\caption{\label{fig:cavity48f} Alternative MAD-EP-CPA configuration (``Design~B'').
(a)~Rather than coupling two $4f$-cavities as in Design~A,  a coupling mirror ($R_\mathrm{c}$) is positioned at the center of a single $4f$-cavity. The left mirror is divided into an input-coupling-part ($R_{1\mathrm{a}}$) illustrated above the optical axis, and a completely reflecting mirror ($R_{1\mathrm{b}}$) depicted below the optical axis. The absorber with transmissivity $T_{\mathrm{a}}$  is placed next to mirror $R_{1\mathrm{b}}$. To correct for refraction aberrations, the left lens's focal length is adjusted to $f'$, and a transparent slab $T_1$, matching $T_\text{a}$ in thickness and real value of refractive index, is placed near $R_{1\mathrm{a}}$.
Exemplary beam paths are marked in red and blue
above and below the optical axis.
(b)~Numerical simulation of the field-of-view reflection spectrum (blue line) in comparison to an ideal quartic behavior (broad yellow line). (c)~Input field intensity pattern.
(d)~Reflected field intensity pattern, demonstrating perfect absorption.} 
\end{figure}
However, Design~A poses challenges in the implementation, as it involves weakly coupling two plane-parallel cavities with altogether four embedded lenses. We thus propose in Fig.~\ref{fig:cavity48f}(a) an alternative ``Design~B'', involving a coupling mirror with reflectivity $R_\mathrm{c}$ at the center of a \textit{single} $4f$-cavity, requiring only two lenses. This simplification requires a division of the left mirror  at the optical axis into an input-coupling segment with reflectivity $R_{1\mathrm{a}}<1$ and a fully reflective mirror $R_{1\mathrm{b}}=1$. The weak absorber is then placed adjacent to the totally reflective mirror $R_{1\mathrm{b}}$. Additionally, to mitigate the additional refraction introduced by the absorber, we adjust the focal length of the left lens to $f'=f-d(n_\text{r}-1/n_\text{r})/2$  (see supplementary \cite[S8]{SM}). Furthermore, to ensure an identical optical path-length above and below the optical axis, we have incorporated a transparent slab, designated $T_1$, having the same thickness and refractive index as the absorber, adjacent to mirror $R_{1\mathrm{b}}$.  

Design~B can be seen as a ``folded-up'' variant of Design~A by identifying analogous round-trip paths. For instance, in Design~A, light undergoes self-imaging round-trips within the right sub-cavity, with a corresponding path in Design~B from the bottom-left to the top-right. A similar parallel can be observed for round-trips in Design A's left sub-cavity, corresponding with the path from top-left to bottom-right in Design~B. In both of these scenarios, a \textit{single reflection} at Design~A's central mirror is substituted by \textit{two transmissions} through Design B's central mirror.
 Also, for the third round-trip-path in Design~A, where light propagates between the farthest mirrors $R_1$ and $R_2$, an analog round-trip-path can be identified in Design~B (see supplementary \cite[S7]{SM} for details). However, it is crucial to recognize that such comparisons between Design A and Design B offer, at best, a qualitative analogy. A comprehensive comparison requires a sound mathematical analysis, considering the infinite sum of all possible round trips and the effects of coupling between the two sub-cavities. Nonetheless, this qualitative comparison hints that the condition for the \textit{transmissivity} $T_\text{c}$ of the central mirror in Design~B may resemble the condition for the \textit{reflectivity} $R_\text{c}$ of the central mirror in Design~A. Using an analogous derivation method as before, the following matrix equation for $\textbf{R}_{\text{\tiny{CPA}}}^{\text{\tiny{(B)}}}$ can be established:
\begin{align}
    \textbf{R}_\text{CPA}^\text{(B)} = \frac{t_\text{a}^2 \, \textbf{T}_\text{2f}^{8}+( 1 + r_\text{1a} t_\text{a}^2  - a ) \, \textbf{T}_\text{2f}^4  + r_\text{1a} \mathbb{1} }{r_\text{1a} t_\text{a}^2 \, \textbf{T}_\text{2f}^{8} + (r_\text{1a} + t_\text{a}^2  -a ) \, \textbf{T}_\text{2f}^{4} + \mathbb{1}},
    \label{eq:RLa}
\end{align}
with the scalar value $a = r_\text{c}^2(t_\text{a}^2+1)(r_\text{1a}^2+1)$. Observe how the structure of Eq.~(\ref{eq:RLa}) resembles the structure of Eq.~(\ref{eq:RCPA_simplified}). Because $\textbf{T}_\text{2f}^4 = \textbf{T}_\text{4f}^2$, the matrix $\textbf{T}_\text{2f}^4$ in Eq.~(\ref{eq:RLa}) results in a quadruple Fourier transformation with a uniform phase shift $e^{2i\phi}$ for all modes, and can thus be represented as a simple diagonal matrix $\textbf{T}_\text{4f}^2=e^{2i\phi} \mathbb{1}$. Analogously, $\textbf{T}_\text{2f}^{8}=e^{4i\phi} \mathbb{1}$. Consequently, $\textbf{R}_{\text{\tiny{CPA}}}^{\text{\tiny{(B)}}}$ becomes a diagonal matrix and can be reduced to a simple scalar equation, which proves the \textit{spatial} degeneracy of Design~B. The \textit{spectral} degeneracy condition can be met when the following EP-conditions are fulfilled (see supplementary \cite[S6]{SM}):
\begin{align}
R_{1\text{a}} &= T_\text{a}^2 \label{eq:R148f} \\
R_\mathrm{c} = 
\left(\frac{1-T_\mathrm{a}}{1+T_\mathrm{a}}\right)^2
\;\; &\Rightarrow \;\;
T_\mathrm{c} = 
\frac{4 T_\mathrm{a}}{(1+T_\mathrm{a})^2}
\label{eq:Rc48f} 
\end{align}
Note the similarity between Eqs. (\ref{eq:R148f}), (\ref{eq:Rc48f}) and (\ref{eq:EP_conditions}).
While the central mirror in Design~A must exhibit \textit{high reflectivity} at the Exceptional Point (EP), in Design~B it requires \textit{high transmissivity}, which is in accordance with our earlier qualitative analysis. The analytical results confirm the anticipated quartic behavior and have been corroborated by numerical computer simulations, as illustrated in Fig.~\ref{fig:cavity48f}. 

\textit{Conclusions.}\,\textemdash\,We have introduced  two CPA designs that combine spectral and spatial degeneracy. Through analytical and numerical models, we have validated the feasibility of this concept and explored its limitations. This new type of CPA is capable of perfectly depositing a light beam in a weak absorber over a wider spectral range as compared to a conventional CPA and regardless of the light beam's spatial wavefront. 

\textit{Acknowledgements}\,\textemdash\,We would like to express our gratitude to Maximilian Prüfer (TU Wien), Jacob Scheuer (Tel Aviv University), and Birgitta Schultze-Bernhardt (TU Graz) for very helpful discussions. The numerical calculations were performed on the Vienna Scientific Cluster (VSC). 


\bibliography{EPMADCPAarx}

\end{document}


\preprint{APS/123-QED}

\title{Supplementary Materials for \\ Coherent Perfect Absorption of Arbitrary Wavefronts\\ at an Exceptional Point}

\author{Helmut Hörner}
\affiliation{Institute for Theoretical Physics, TU Wien, Wiedner Hauptstra\ss e 8-10/136, A-1040 Vienna, Austria}

\author{Lena Wild}
\affiliation{Institute for Theoretical Physics, TU Wien, Wiedner Hauptstra\ss e 8-10/136, A-1040 Vienna, Austria}

\author{Yevgeny Slobodkin}
\affiliation{Institute of Applied Physics, The Hebrew University of Jerusalem, Bergman 206, 9190401, Jerusalem, Israel}

\author{Gil Weinberg}
\affiliation{Institute of Applied Physics, The Hebrew University of Jerusalem, Bergman 206, 9190401, Jerusalem, Israel}

\author{Ori Katz}
\affiliation{Institute of Applied Physics, The Hebrew University of Jerusalem, Bergman 206, 9190401, Jerusalem, Israel}

\author{Stefan Rotter} 
\affiliation{Institute for Theoretical Physics, TU Wien, Wiedner Hauptstra\ss e 8-10/136, A-1040 Vienna, Austria}

\date{\today}

\maketitle
\newpage
\tableofcontents
\newpage
\section{S1. Deriving the reflection matrix for an MAD-EP-CPA with two cavities}

\noindent Here, we derive an expression for the reflection matrix $\textbf{R}_\text{L}$ for an MAD-EP-CPA with two cavities (Eq.~(1) in the main text). To do so, we employ a scalar optics model for polarized light, considering a large, yet finite, number of spatial modes. Let $\textbf{R}_1$, $\textbf{R}_2$, and $\textbf{R}_c$ represent the reflection matrices, and $\textbf{T}_1$, $\textbf{T}_2$, $\textbf{T}_\text{c}$, and $\textbf{T}_\text{a}$  the transmission matrices for the left, right, central mirror, and the absorber, respectively. Additionally, let $\textbf{T}_\text{4f}$ represent the transmission matrix for propagation through a $4f$-system, comprising two lenses in a telescopic configuration. $\textbf{T}_\text{4f}$ acts as a double Fourier transform, organizing the development coefficients of the $\textbf{k}_\text{xy}$ vectors to produce an image that is flipped in the spatial domain. Moreover, it applies a uniform phase $e^{i \phi}$ to all modes. From these individual transmission matrices, scattering matrices for each optical component and optical propagation can be constructed. These scattering matrices define the relationship between the left-and-right outgoing field amplitude vectors $\mathbf{a}_\mathrm{out}$ in relation to the left-and-right incident amplitude vectors, $\mathbf{a}_\mathrm{in}$. For this particular derivation, we adopt the following convention for any scattering matrix  $\mathbf{S}$:
\begin{equation}
    \left( \begin{array}{ll} \mathbf{a}^{\mathrm{left}}_{\mathrm{out}} \\ 
    \mathbf{a}^{\mathrm{right}}_{\mathrm{out}}  \end{array} \right) = 
    \mathbf{S} 
    \left( \begin{array}{ll} \mathbf{a}^{\mathrm{left}}_{\mathrm{in}} \\ 
    \mathbf{a}^{\mathrm{right}}_{\mathrm{in}}  \end{array} \right) = 
    \left( \begin{array}{cc} \mathbf{R}_\mathrm{L} & \mathbf{T}_\mathrm{RL} \\ \mathbf{T}_\mathrm{LR} & \mathbf{R}_\mathrm{R} \end{array} \right)
    \left( \begin{array}{ll} \mathbf{a}^{\mathrm{left}}_{\mathrm{in}} \\ 
    \mathbf{a}^{\mathrm{right}}_{\mathrm{in}}  \end{array} \right) \label{eq:smatrix}
\end{equation}
The scattering matrix $\mathbf{S}$ is a block matrix composed of four sub-matrices: $\mathbf{R}_\mathrm{L}$  and $\mathbf{R}_\mathrm{R}$ are the reflection matrices for light incident from the left and right, respectively, while $\mathbf{T}_\mathrm{LR}$ and $\mathbf{T}_\mathrm{RL}$ are the transmission matrices for light passing from left-to-right and right-to-left, respectively. Although these matrices inherently contain complex values to account for phase shifts, the analysis can be simplified. As detailed, for example, in [1,~p.~406], it is customary in single-mode problems (without loss of generality) to represent a scattering matrix using only real numbers, $r$ and $t$, by adopting the following convention:
\begin{eqnarray}
\mathbf{S}=\begin{pmatrix}
r_\mathrm{L} & t_\mathrm{RL} \\
 t_\mathrm{LR} & r_\mathrm{R}
\end{pmatrix} =
\begin{pmatrix}
r & t \\ t & -r
\end{pmatrix} \;\;\; r,t \in \left[ 0, 1\right] \label{eq:defS}
\end{eqnarray}
This convention ensures the energy-conserving unitarity of the scattering matrix, so that $\mathbf{S}\mathbf{S}^\dagger=\openone$, where $\openone$ represents the identity matrix.  Assuming that the same reflection and transmission coefficients apply to all modes for the mirrors and the absorber, the corresponding reflection and transmission matrices $\mathbf{R}$ and $\mathbf{T}$ become diagonal matrices. Hence, convention (\ref{eq:defS}) can readily be generalized to our block-matrix-style scattering matrices:
\begin{eqnarray}
\mathbf{S}=\begin{pmatrix}
\mathbf{R}_\mathrm{L} & \mathbf{T}_\mathrm{RL} \\
 \mathbf{T}_\mathrm{LR} & \mathbf{R}_\mathrm{R}
\end{pmatrix} =
\begin{pmatrix}
\mathbf{R} & \mathbf{T} \\ \mathbf{T} & -\mathbf{R}
\end{pmatrix} \;\;\; \mathbf{R}, \mathbf{T} \in \mathbb{R}^{n \times n} \label{eq:defS2}
\end{eqnarray}
Given this convention, the scattering matrices for the left, central, and right mirrors can be expressed in the following manner:
\begin{equation} 
\mathbf{S}_1 =   \begin{pmatrix} \mathbf{R}_1 & \mathbf{T}_1 \\ \mathbf{T}_1 & -\mathbf{R}_1 \end{pmatrix} \;\;\;\;\;
\mathbf{S}_\mathrm{c} =   \begin{pmatrix} \mathbf{R}_\mathrm{c} & \mathbf{T}_\mathrm{c} \\ \mathbf{T}_\mathrm{c} & -\mathbf{R}_\mathrm{c} \end{pmatrix}  \;\;\;\;\;
\mathbf{S}_2 =   \begin{pmatrix} \mathbf{R}_2 & \mathbf{T}_2 \\ \mathbf{T}_2 & -\mathbf{R}_2 \end{pmatrix} \label{eq:defS1ScS2}
\end{equation}

\noindent The single-trip propagation in either direction between the mirrors over the distance $4f$ and the absorption in the absorber is represented by the following scattering matrices:
\begin{equation} 
\mathbf{S}_\mathrm{4f} =   \begin{pmatrix} \mathbb{0} & \textbf{T}_\text{4f} \\ \textbf{T}_\text{4f} & \mathbb{0} \end{pmatrix}  \;\;\;\;\;
\mathbf{S}_\mathrm{a} =   \begin{pmatrix} \mathbb{0} & \mathbf{T}_\mathrm{a} \\ \mathbf{T}_\mathrm{a} & \mathbb{0} \end{pmatrix} \label{eq:defSpropSa}
\end{equation}
The next step is to convert all scattering matrices into transfer matrices. These transfer matrices relate the incident and outgoing amplitude vectors on the left side to their respective vectors on the right side. For this conversion, we adhere to the following convention for the transfer matrices:
\begin{equation}
    \left( \begin{array}{ll} \mathbf{a}^{\mathrm{left}}_{\mathrm{out}} \\ 
    \mathbf{a}^{\mathrm{left}}_{\mathrm{in}}  \end{array} \right) = 
    \mathbf{M} 
    \left( \begin{array}{ll} \mathbf{a}^{\mathrm{right}}_{\mathrm{in}} \\ 
    \mathbf{a}^{\mathrm{right}}_{\mathrm{out}}  \end{array} \right) \label{eq:defM}
\end{equation}
In order to transform the above-defined scattering matrices into their corresponding transfer matrices, and subsequently back-convert the final transfer matrix $\mathbf{M}_\mathrm{CPA}^\text{(A)}$ for the entire CPA (Design~A) into the scattering matrix $\mathbf{S}_\mathrm{CPA}^\text{(A)}$, we employ the generalized conversions outlined in [2]:
\begin{eqnarray}
\mathbf{M}(\mathbf{S})=
\begin{pmatrix}
\mathbf{S}_{12}-\mathbf{S}_{11}  \mathbf{S}_{21}^{-1} \mathbf{S}_{22} \;\;& 
\mathbf{S}_{11} \mathbf{S}_{21}^{-1} \\ 
-\mathbf{S}_{21}^{-1} \mathbf{S}_{22} &
 \mathbf{S}_{21}^{-1}
\end{pmatrix} \;\;
\mathbf{S}(\mathbf{M})=
\begin{pmatrix}
\mathbf{M}_{12} \mathbf{M}_{22}^{-1} \;\; &
\mathbf{M}_{11}-\mathbf{M}_{12}  \mathbf{M}_{22}^{-1} \mathbf{M}_{21} \\
 \mathbf{M}_{22}^{-1} &
-\mathbf{M}_{22}^{-1} \mathbf{M}_{21}
\end{pmatrix}
\label{eq:defSMconversion2}
\end{eqnarray}
In (\ref{eq:defSMconversion2}), $\mathbf{S}_{11}$ is the top-left-quadrant sub-matrix, $\mathbf{S}_{12}$ is the top-right-quadrant sub-matrix, $\mathbf{S}_{21}$ is the bottom-left-quadrant sub-matrix, and $\mathbf{S}_{22}$ is the bottom-right-quadrant sub-matrix of the scattering matrix $\mathbf{S}$. Analogously, $\mathbf{M}_{11}$, $\mathbf{M}_{12}$, $\mathbf{M}_{21}$, and $\mathbf{M}_{22}$ represent the respective quadrants of the transfer matrix $\mathbf{M}$. By employing the $\mathbf{M}(\mathbf{S})$ conversion, we are equipped to transform all the scattering matrices from Eq.~(\ref{eq:defS1ScS2}) and Eq.~(\ref{eq:defSpropSa}) into their respective transfer matrices. Subsequently, this facilitates the straightforward calculation of the overall transfer matrix for the CPA, which can be expressed as:
\begin{equation}
    \mathbf{M}_\mathrm{CPA}^\text{(A)} = \mathbf{M}_1 \mathbf{M}_\mathrm{4f} \mathbf{M}_\mathrm{c}  \mathbf{M}_\mathrm{4f} \mathbf{M}_\mathrm{a}
    \mathbf{M}_2 =
    \mathbf{M}(\mathbf{S}_1) \,
    \mathbf{M}(\mathbf{S}_\mathrm{4f}) \,
    \mathbf{M}(\mathbf{S}_\mathrm{c}) \, 
    \mathbf{M}(\mathbf{S}_\mathrm{4f}) \,
    \mathbf{M}(\mathbf{S}_\mathrm{a}) \,
    \mathbf{M}(\mathbf{S}_2)
    \label{eq:Mchain}
\end{equation}
Considering that all the $\mathbf{M}$-matrices mentioned are $2 \times 2$ block matrices, with each quadrant being a matrix in its own right, the following multiplication rule for $2 \times 2$ block matrices $\mathbf{X}$ and $\mathbf{Y}$ must be applied for calculating expression (\ref{eq:Mchain}) and any subsequent equivalent expressions:
\begin{equation}
    \mathbf{X} \mathbf{Y} = 
    \begin{pmatrix}
    \mathbf{X}_{11} \mathbf{Y}_{11} +
    \mathbf{X}_{12} \mathbf{Y}_{21} \;\;\;&
    \mathbf{X}_{11} \mathbf{Y}_{12} +
    \mathbf{X}_{12} \mathbf{Y}_{22} \\
    \mathbf{X}_{21} \mathbf{Y}_{11} +
    \mathbf{X}_{22} \mathbf{Y}_{21} \;\;\;&
    \mathbf{X}_{21} \mathbf{Y}_{12} +
    \mathbf{X}_{22} \mathbf{Y}_{22}
    \end{pmatrix}
    \label{eq:BMmult2}
\end{equation}
Using the reverse conversion $\mathbf{S}(\mathbf{M})$ from Eq. (\ref{eq:defSMconversion2}), we can subsequently obtain the scattering matrix representing the entirety of the CPA system:
\begin{equation}
    \mathbf{S}_\mathrm{CPA}^\text{(A)} = \mathbf{S}(\mathbf{M}_\mathrm{CPA}^\text{(A)})
\end{equation}
Our main interest is the top-left quadrant of the scattering matrix $\mathbf{S}_\mathrm{CPA}^\text{(A)}$, which corresponds to the CPA's reflection matrix for Design~A, denoted as $ \mathbf{R}_\text{CPA}^\text{(A)}$:

\begin{equation}
    \textbf{R}_\text{CPA}^\text{(A)} = \frac{(\textbf{R}_1  \textbf{A} 
    + \textbf{B} \textbf{E} ) \, \textbf{D} +
    ( \textbf{B} \textbf{C} +
    \textbf{R}_1 \textbf{A} \, \textbf{R}_\text{c}
    ) \, \textbf{G}}{( \textbf{F} \textbf{E} +
    \textbf{A})\, \textbf{D} + (\textbf{A} \textbf{R}_\text{c} +
    \textbf{F} \textbf{C} ) \, \textbf{G} } \label{eq:RCPA}
\end{equation}
\noindent with the following substitutions:  
\begin{align*}
    \textbf{A} &= \left( \textbf{T}_\text{c} \textbf{T}_\text{4f} \textbf{T}_1\right)^{-1} \\
    \textbf{B} &= \textbf{T}_1 + \textbf{R}_1 \textbf{T}_1^{-1} \textbf{R}_1 \\
    \textbf{C} &= \textbf{T}_\text{4f} (\textbf{T}_\text{c} + \textbf{R}_\text{c} \textbf{T}_\textbf{c}^{-1} \textbf{R}_\text{c}) \\
    \textbf{D} &= ( \textbf{T}_\text{a} \textbf{T}_\text{4f} )^{-1} \\
    \textbf{E} &= \textbf{T}_\text{4f} \, \textbf{R}_\text{c} \textbf{T}_\text{c}^{-1} \\
    \textbf{F} &= \textbf{T}_1^{-1} \textbf{R}_1 \\
    \textbf{G} &= \textbf{T}_\text{4f} \, \textbf{T}_\text{a} \textbf{R}_\text{2}
\end{align*}
Assuming that the same reflection and transmission coefficients apply to all modes at the mirrors and the absorber, all matrices except $ \textbf{T}_\text{4f}$ become diagonal matrices:
\begin{align}
    \textbf{R}_1 = r_1 \mathbb{1}, \;\;
    \textbf{R}_2 = r_2 \mathbb{1}, \;\;
    \textbf{R}_\text{c} = r_\text{c} \mathbb{1}, \;\;
    \textbf{T}_1 = t_1 \mathbb{1}, \;\;
    \textbf{T}_2 = t_2 \mathbb{1}, \;\;
    \textbf{T}_\text{c} = t_\text{c} \mathbb{1}, \;\;
    \textbf{T}_\text{a} = t_\text{a} \mathbb{1}
    \label{eq:diagsubst}
\end{align}
Thus, Eq. (\ref{eq:RCPA}) can be significantly simplified to
\begin{align}
    \textbf{R}_\text{CPA}^\text{(A)} = \frac{t_\text{a}^2  (r_1^2 + t_1^2) (r_\text{c}^2 + t_\text{c}^2) r_2
    \, \textbf{T}_\text{4f}^4 +r_\text{c}   (r_1^2 + t_1^2 + r_1 r_2 t_\text{a}^2)  \,\textbf{T}_\text{4f}^2   + r_1 \mathbb{1}}{r_1 r_2 t_\text{a}^2 (r_\text{c}^2 + t_\text{c}^2) \, \textbf{T}_\text{4f}^4 + r_\text{c}(r_1+ r_2 t_\text{a}^2)  \, \textbf{T}_\text{4f}^2 + \mathbb{1}} \label{eq:RCPA_simplified1}
\end{align}
Considering the energy conservation relations $ r_1^2 + t_1^2 = 1$ and $r_c^2 + t_c^2 = 1$, and assuming $r_2 = 1$ (indicating a totally reflective mirror on the right side), the equation simplifies to:
\begin{align}
    \textbf{R}_\text{CPA}^\text{(A)} = \frac{t_\text{a} ^2 \, \textbf{T}_\text{4f}^4 +r_\text{c}   (1+r_1 t_\text{a}^2)  \,\textbf{T}_\text{4f}^2   + r_1 \mathbb{1}}{r_1 t_\text{a}^2 \, \textbf{T}_\text{4f}^4 + r_\text{c}(r_1+t_\text{a}^2)  \, \textbf{T}_\text{4f}^2 + \mathbb{1}} \label{eq:RCPA_simplified2}
\end{align}
This corresponds to Eq.~(1) in the main paper. The matrix $\textbf{T}_\text{4f}^2$ results in a quadruple Fourier transformation with a uniform phase shift $e^{2i\phi}$ for all modes and can thus be represented as a simple diagonal matrix $\textbf{T}_\text{4f}^2=e^{2i\phi} \mathbb{1}$. Analogously, $\textbf{T}_\text{4f}^4=e^{4i\phi} \mathbb{1}$. Consequently, $\textbf{R}_\text{CPA}^\text{(A)}$ becomes a diagonal matrix with identical entries $r_\text{CPA}^\text{(A)}(\omega)$ along the diagonal and can be reduced to a simple scalar equation, which proves the spatial degeneracy of the system.
\begin{equation}
    r_\mathrm{CPA}^\text{(A)} = 
    \frac{t_\mathrm{a}^2 e^{4 i \phi} +
    r_\mathrm{c} \left(1 + r_1 t_\mathrm{a}^2\right) e^{2 i \phi} + r_1}
    {r_1 t_\mathrm{a}^2   e^{4 i \phi}   +
    r_\mathrm{c}  \left(r_1+ t_\mathrm{a}^2\right) e^{2 i \phi} 
    + 1}. \label{eq:rLEPCPA}
\end{equation}
In Eq.~(\ref{eq:rLEPCPA}), $r_1$ is the reflection-coefficient of the left mirror, $r_\text{c}$ is the reflection-coefficient of the central mirror, and $t_\mathrm{a}$ is the one-way transmission coefficient of the weak absorber. For an EP-CPA, we first need to set the expression in Eq.~(\ref{eq:rLEPCPA}) to zero for total absorption. Subsequently, we  solve for $\phi$, to find the parameters where the two distinct solutions merge to an Exceptional Point.

\section{S2. Deriving the CPA conditions for an Exceptional Point CPA with two cavities}

\noindent In this section,  the derivation of the CPA conditions for an EP-CPA with two cavities are explained in detail. For total absorption, we set the expression in Eq.~(\ref{eq:rLEPCPA}) to zero, and then solve for $\phi$:
\begin{eqnarray}
&&\phi=\pi m - \frac{i}{2}\log{\left(
-\frac{r_\mathrm{c}  +
r_1 r_\mathrm{c} t_\mathrm{a}^2}
{2 t_\mathrm{a}^2} 
 \pm \frac{\sqrt{\left( r_\mathrm{c} + r_1 r_\mathrm{c} t_\mathrm{a}^2 \right)^2 -
4 r_1 t_\mathrm{a}^2}}
{2 t_\mathrm{a}^2} \right)}
\label{eq:three} 
\;\;\; \mathrm{with}\;\;\; m \in \mathbb{N} 
\end{eqnarray}
At the exceptional point, the two distinct solutions for $\phi$ in Eq.~(\ref{eq:three}) coincide. Consequently, the expression beneath the square root must become zero.
\begin{eqnarray}
 \left( r_\mathrm{c} + r_1 r_\mathrm{c} t_\mathrm{a}^2 \right)^2 -
4 r_1 t_\mathrm{a}^2 = 0
\label{eq:four} 
\end{eqnarray}
for $r_\mathrm{c}$, we find
\begin{eqnarray}
r_\mathrm{c} = \frac{2 \sqrt{r_1} t_\mathrm{a}}{1 + r_1 t_\mathrm{a}^2}
\label{eq:five} 
\end{eqnarray}
By substituting this result into Eq.~(\ref{eq:rLEPCPA}), and setting the equation to zero again, we can solve for the transmission coefficient $t_\mathrm{a}$. This leads to the derivation of the relationship $t_\mathrm{a} = -  e^{-2 i \phi} \sqrt{r_1}$. We require $t_\mathrm{a} \in \mathbb{R}_+$. This is the case when $e^{-2 i \phi} = -1$, which is equivalent to $\phi_m = \frac{\pi +2 \pi  m}{2}$ ($m \in \mathbb{Z}$). These particular values of $\phi_m$ correspond to the resonance wave numbers $k_m = \frac{\phi_m}{4f}$ that yield perfect absorption. Assuming this condition,  we proceed to write
\begin{eqnarray}
t_\mathrm{a} = \sqrt{r_1} 
\label{eq:six} 
\end{eqnarray}
Plugging this into Eq.~(\ref{eq:five}) yields
\begin{eqnarray}
r_\mathrm{c} = \frac{2r_1}{1+r_1^2}
\label{eq:rcEPCPA} 
\end{eqnarray}
By squaring both sides of Eq.~(\ref{eq:six}) and substituting $t_\mathrm{a}^2 \rightarrow T_\mathrm{a}$ and $r_1 \rightarrow \sqrt{R_1}$, we get
\begin{eqnarray}
T_\mathrm{a} = \sqrt{R_1}
\;\;\Leftrightarrow\;\;
R_1 = T_\mathrm{a}^2
\end{eqnarray}
Finally, squaring both sides of Eq.~(\ref{eq:rcEPCPA}), and substituting $r_\mathrm{c}^2 \rightarrow R_\mathrm{c}$, $r_1^2 \rightarrow R_1$ leads to
\begin{eqnarray}
R_\mathrm{c} = \frac{4 R_1}{\left(1+R_1 \right)^2}
\label{eq:Rc8}
\end{eqnarray}
This proves the EP-conditions in Eq. (2) of the main text.\\

\section{S3. Alternative Derivation of CPA Conditions  for 2-Cavity-EP-CPA and Quartic Behavior Proof}

\noindent In this section we demonstrate an alternative approach for deriving the CPA conditions for an EP-CPA with 2 cavities. In contrast to the previous method, this approach can be generalized for CPAs with more than two cavities. For perfect absorption, the condition $R_\text{CPA}^\text{(A)}(\phi) = |r_\mathrm{CPA}^\text{(A)}(\phi)|^2  = 0$ must be satisfied. Calculating $R_\mathrm{CPA}^\text{(A)}$ yields
\begin{equation}
    R_\mathrm{CPA}^\text{(A)}(\phi) = 
    \frac{r_1^2 r_\mathrm{c}^2 t_\mathrm{a}^4+r_1^2+2 r_1 r_\mathrm{c}^2 t_\mathrm{a}^2+2 r_\mathrm{c} \left(r_1+t_\mathrm{a}^2\right) \left(r_1 t_\mathrm{a}^2+1\right) \cos (2 \phi )+2 r_1 t_\mathrm{a}^2 \cos (4 \phi )+r_\mathrm{c}^2+t_\mathrm{a}^4}{t_\mathrm{a}^4 \left(r_1^2+r_\mathrm{c}^2\right)+r_1^2 r_\mathrm{c}^2+2 r_1 r_\mathrm{c}^2 t_\mathrm{a}^2+2 r_\mathrm{c} \left(r_1+t_\mathrm{a}^2\right) \left(r_1 t_\mathrm{a}^2+1\right) \cos (2 \phi )+2 r_1 t_\mathrm{a}^2 \cos (4 \phi )+1}
    \label{eq:RLofPhi}
\end{equation}

\begin{figure}[h]
\includegraphics[width=10cm]{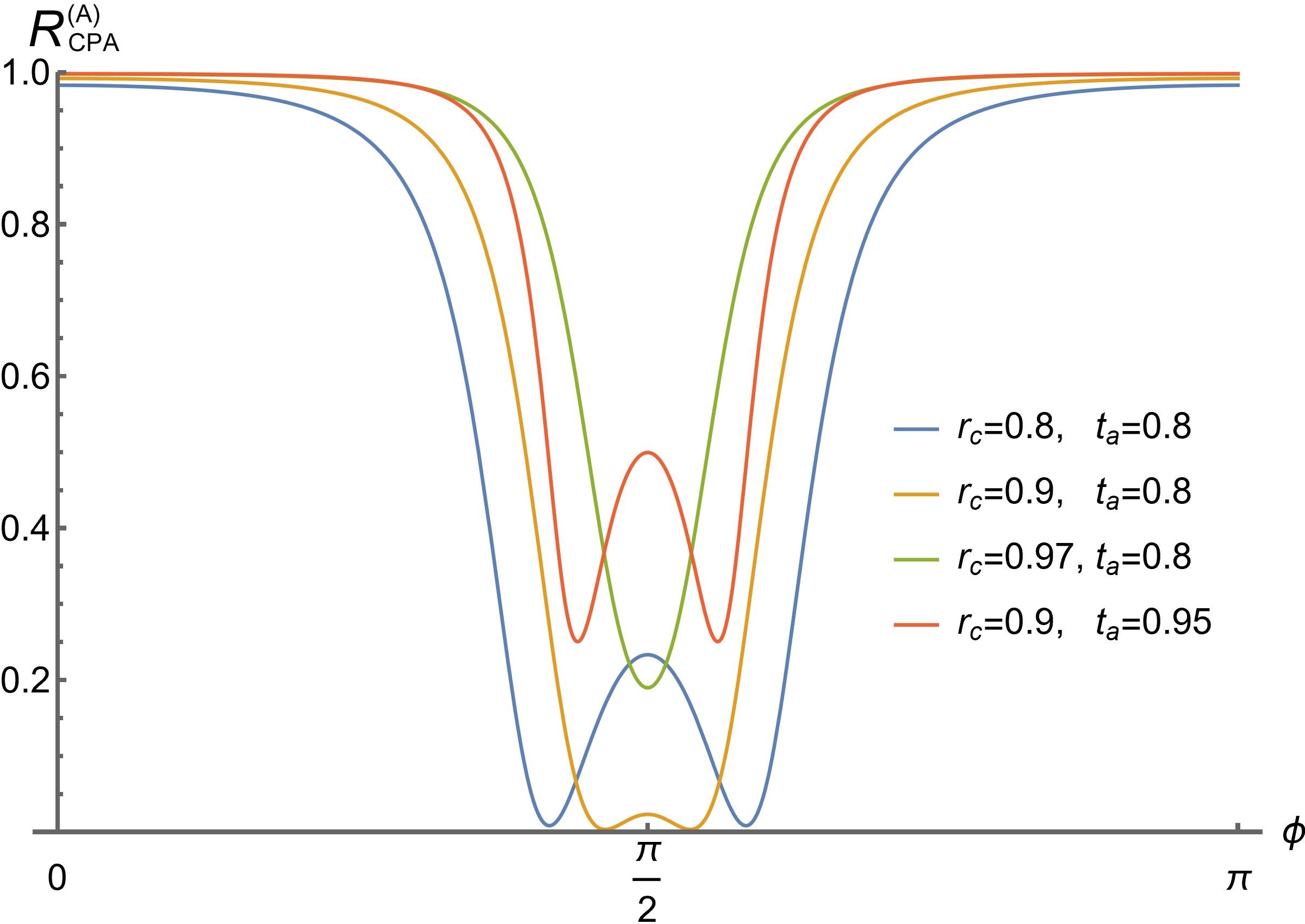}
\caption{\label{fig:deriv_EP_CPA2a} A plot of $R_\mathrm{CPA}^\text{(A)}(\phi)$ with $r_1=0.7$ and some arbitary values for $r_\mathrm{c}$ and $ t_\mathrm{a}$ shows that we can expect to find perfect absorption at $\phi=\frac{\pi}{2}$, provided we find the right values for $r_\text{c}$ and $t_\text{a}$.} 
\end{figure}

\noindent Plotting $R_\mathrm{CPA}^\text{(A)}(\phi)$ with a fixed value for $r_1$ and different arbitrary values for $r_\mathrm{c}$ and $ t_\mathrm{a}$, as depicted in Fig.~\ref{fig:deriv_EP_CPA2a},  it becomes evident that perfect absorption can be anticipated at $\phi=\frac{\pi}{2}$, provided we can find the appropriate values for $r_\mathrm{c}$ and $ t_\mathrm{a}$.
However, our objective is not solely to ensure that $R_\mathrm{CPA}^\text{(A)}(\phi)$ equates to zero at $\phi=\frac{\pi}{2}$; we also want to observe the broadening of the absorption dip caused by the exceptional point. This is the case when $R_\mathrm{CPA}^\text{(A)}(\phi)$ exhibits a saddle point at $\phi=\frac{\pi}{2}$. Mathematically, this is equivalent to asserting that the second derivative of $R_\mathrm{CPA}^\text{(A)}(\phi)$ with respect to $\phi$ also vanishes at $\phi=\frac{\pi}{2}$. This leads to the following system of equations:
\begin{eqnarray}
\left. R_\mathrm{CPA}^\text{(A)}\left(\phi\right)\right\vert_{\phi=\frac{\pi}{2}} &= 0 \\
\partial_\phi^2 \left. R_\mathrm{CPA}^\text{(A)}\left(\phi\right)\right\vert_{\phi=\frac{\pi}{2}} &= 0
\end{eqnarray}
Solving this system of equations yields the solutions $t_\mathrm{a}=\sqrt{r_1}$ and $r_\mathrm{c} = \frac{2r_1}{1+r_1^2}$. These solutions are consistent with the conditions (\ref{eq:six}) and (\ref{eq:rcEPCPA}) derived previously. By incorporating  (\ref{eq:six}) and (\ref{eq:rcEPCPA}) into Eq. (\ref{eq:RLofPhi}), and substituting $r_1^2 = R_1$, we obtain the following expression: 
\begin{equation}
    R_\mathrm{CPA}^\text{(A)}(\phi) = 
    \frac{16\left(1 + R_1 \right)^2 \cos^4(\phi)}{1 + 2 R_1 + 18 R_1^2 + 2 R_1^3 + R_1^4 + 8 R_1 \left(1+R_1 \right)^2 \cos(2 \phi) + 2 R_1 \left(1 + R_1 \right)^2 \cos(4 \phi)} \label{eq:RCPAAopt}
\end{equation}
Developing this expression around the resonance points $\phi_m = \frac{2m+1}{2} \pi$ reveals the anticipated quartic behavior of  $R_\mathrm{CPA}^\text{(A)}$:
\begin{equation}
    R_\mathrm{CPA}^\text{(A)}(\phi) = \frac{16 R_1(R_1+1)^2}{(R_1-1)^4} \left(\phi - \frac{2m+1}{2} \pi \right)^4 + \mathcal{O}^6\left(\phi - \frac{2m+1}{2} \pi \right)^6
\end{equation}

\newpage
\section{S4. Critical Parameters}
\begin{figure}
\includegraphics[width=14cm]{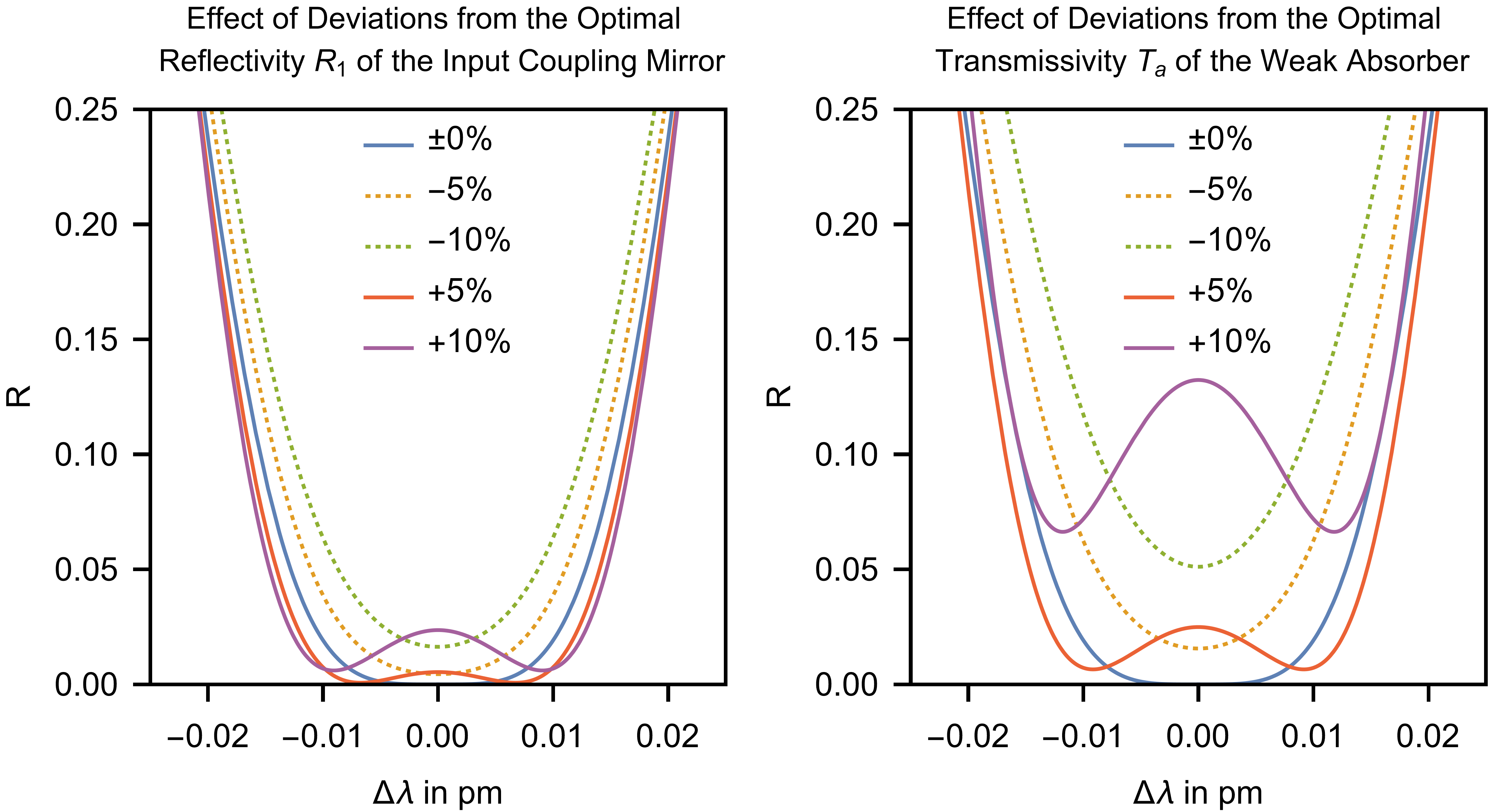}
\caption{\label{fig:parameters-insensitive} (a) Deviations in the reflectivity $R_1$ of the (left) input coupling mirror have a minimal impact on the effectiveness of the EP-CPA. Similarly, variations in the reflectivity $R_2$ of the right totally reflective mirror have negligible effects. (b)~The EP-CPA also demonstrates low sensitivity to variations in the transmissivity $T_\mathrm{a}$ of the absorber. (Assumed wavelength $\lambda_0=633 \mathrm{nm}$)}
\end{figure}

\begin{figure}
\includegraphics[width=15.54cm]{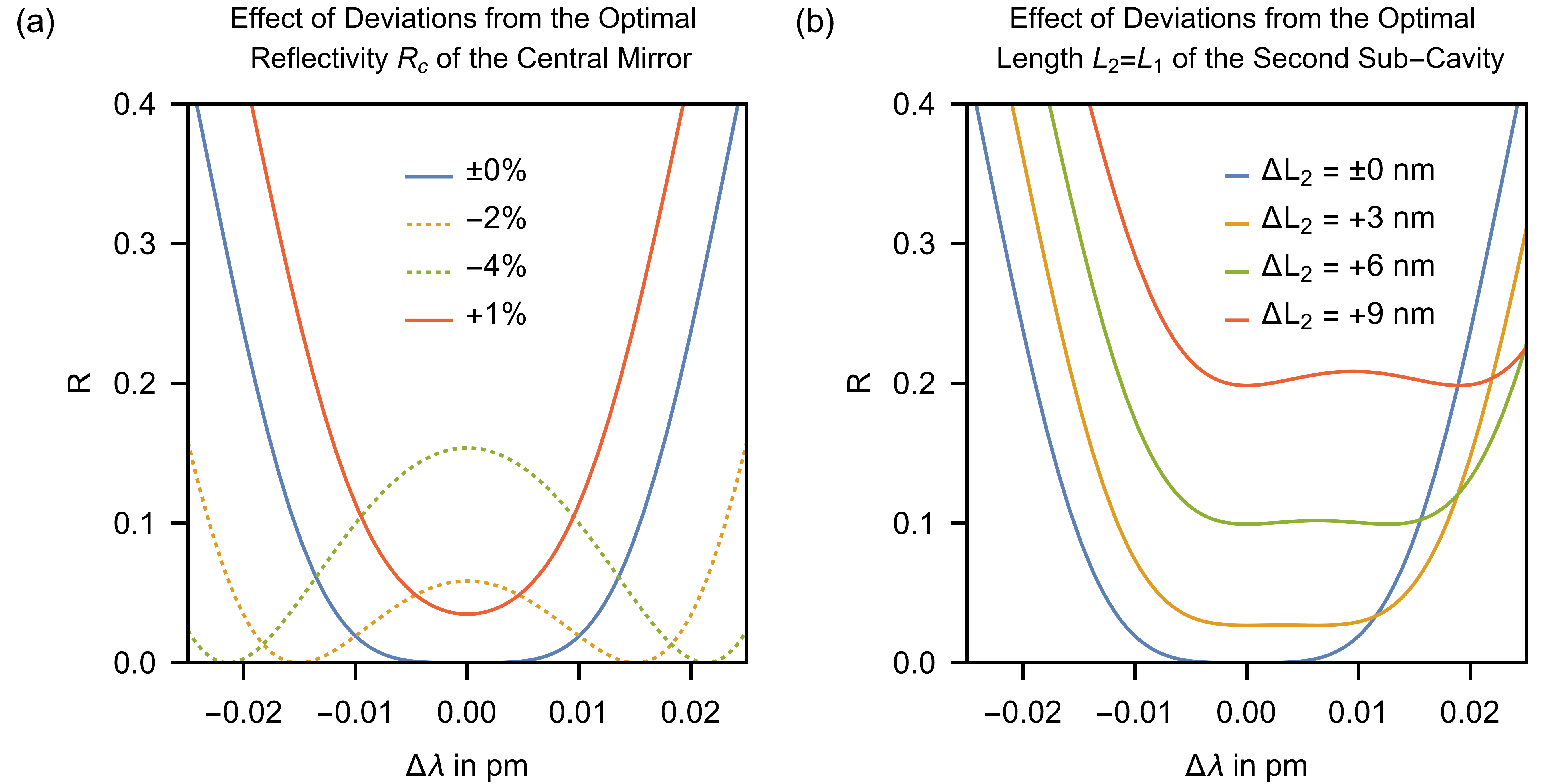}
\caption{\label{fig:parameters-sensitive} (a) Even small deviations in the reflectivity $R_\mathrm{c}$  of the central coupling mirror can significantly impact the effectiveness of the EP-CPA. (b) This is even more evident for minor deviations in the length of the second cavity. (Calculation performed for a reference wavelength $\lambda_0=633 \mathrm{nm}$.)} 
\end{figure}

\noindent For practical implementations, it is insightful to investigate the impact of minor alterations in the system parameters on the efficacy of the MAD-EP-CPA depicted in Fig.~1(d) of the main paper. Some deviations~--~such as the influence of unintended lens reflections or a slight tilt in one of the mirrors -- necessitate a numerical simulation of the actual system for accurate assessment. Nevertheless, to analyze the effects stemming from deviations in the reflectivities of the mirrors or the transmissivity of the absorber, the straightforward analytical solution provided by Eq.~(\ref{eq:RLofPhi}) proves to be adequately sufficient. As is evident from Fig.~\ref{fig:parameters-insensitive}, these calculations reveal that even deviations in the magnitude of a few percent in the mirror reflectivity $R_1$, or in the right absorber's transmissivity $T_\mathrm{a}$ do not substantially compromise the performance of the Exceptional Point CPA. Contrastingly, deviations in the reflectivity $R_\mathrm{c}$ of the central coupling mirror yield significantly different results. As illustrated in Fig.~\ref{fig:parameters-sensitive}(a), even a mere one percent deviation in the central mirror’s reflectivity $R_\mathrm{c}$ can substantially impair the performance of the EP-CPA. The precise alignment of the two cavity lengths is even more crucial. Fig.~\ref{fig:parameters-sensitive}(b) demonstrates that deviations in the cavity lengths equivalent to less than one percent of the wavelength are already consequential.

\begin{figure}[ht]
\includegraphics[width=11cm]{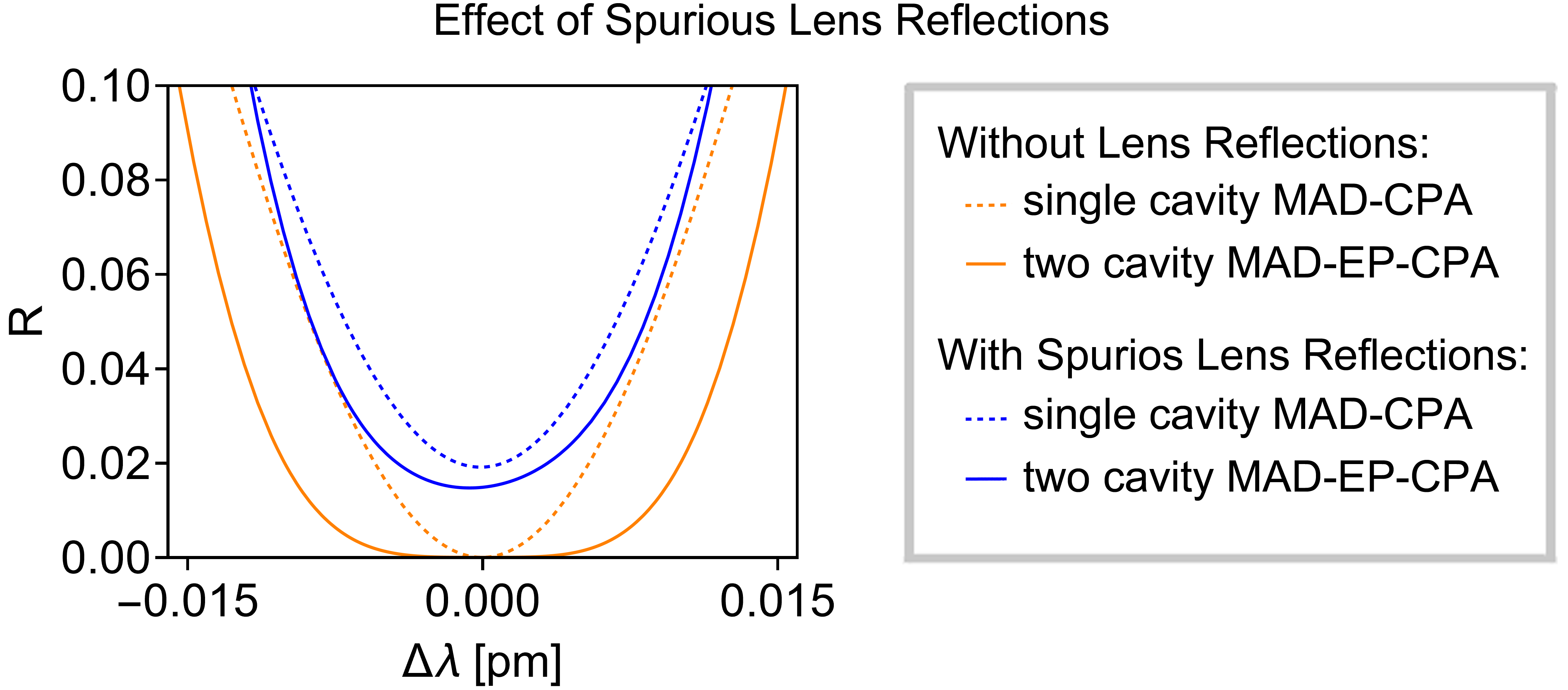}
\caption{\label{fig:lens-refl} Comparison of the field-of-view reflection spectrum of a MAD-EP-CPA with and without spurious lens reflections, using a Scalar Fourier Optics computer simulation. The dotted lines show a single-cavity MAD-CPA for comparison. Note that the assumed residual reflection of  $R=0.13 \%$ per lens-facet affects both the maximal absorption and the EP-broadening-effect (blue lines). Calculation were performed for a reference wavelength $\lambda_0=633 \mathrm{nm}$ and lenses with $f=75\mathrm{mm}$ focal length.} 
\end{figure}

\noindent In reference [3], it was demonstrated that in an actual experiment involving a single-cavity MAD-CPA, the weak spurious reflections originating from the facets of the cavity lenses (with a reflectivity $R=0.13 \%$ per surface) played a significant role in diminishing the CPA effect. Given this, it is anticipated that these spurious reflections would have an even more pronounced impact on the MAD-EP-CPA, as it incorporates four lenses. To accurately model and understand the effects of these spurious lens reflections,  the simple analytical approach utilized thus far is deemed inadequate. Consequently, we resort to the numerical Scalar Fourier Optics computer simulation, a method that has previously demonstrated excellent concordance with experimental measurements, as evidenced in [3]. Fig.~\ref{fig:lens-refl} presents the outcome of such a simulation. In this simulation, a random speckle field consisting of 160 modes of equal amplitude and random phase in $k$-space serves as input to compute the power-reflection spectrum of the MAD-EP-CPA in comparison to a simple, single-cavity MAD-CPA. It illustrates the detrimental impact of the assumed spurious lens reflections, with a reflectivity $R=0.13 \%$ per lens-facet, on the performance of the MAD-EP-CPA. As a result of these spurious reflections, the power reflection over the whole field-of-view does not dip below approximately $2 \%$, and although the EP broadening-effect remains observable, it is notably less prominent compared to the simulation that omits spurious lens reflections. 

\section{S5. Higher-Order Exceptional Point CPAs}
\begin{figure}[ht]
\includegraphics[height=4cm]{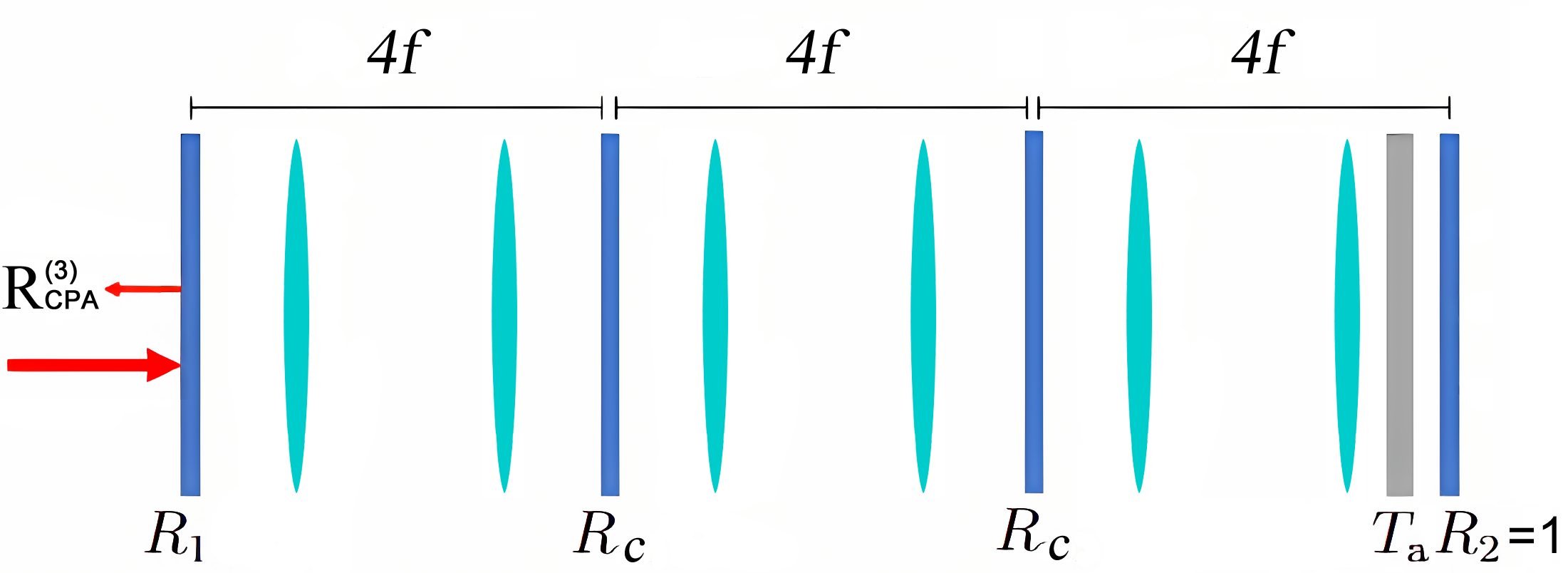}
\caption{\label{fig:EPEP-CPA} Creating an MAD-EP-CPA by critically coupling three $4f$-cavities with equal optical lengths $4f$ and placing a weak absorber in the rightmost cavity adjacent to the totally reflective mirror. With all parameters finely tuned to reach a real-frequency exceptional point, the absorption spectrum exhibits an even broader dip compared to the two-cavity MAD-EP-CPA. }
\end{figure}

\noindent It is possible to extend the concept depicted in Fig.~1(b) and Fig.~1(d) of the main paper by critically coupling more than two sub-cavities. For instance, Fig.~\ref{fig:EPEP-CPA} illustrates a single-mode EP-CPA composed of three cavities, designed to absorb single-mode light waves incident from the left. This configuration can be mathematically represented using the following transfer matrix:
\begin{equation}
    \mathbf{M}_\mathrm{CPA}^\text{(3)} = \mathbf{M}_1 \mathbf{M}_\mathrm{4f} \mathbf{M}_\mathrm{c} \mathbf{M}_\mathrm{4f} \mathbf{M}_\mathrm{c}  \mathbf{M}_\mathrm{4f} \mathbf{M}_\mathrm{a} \mathbf{M}_2
\end{equation}

\noindent Observe that in Fig.~\ref{fig:EPEP-CPA} the two coupling mirrors share the same reflectivity, denoted as $R_\mathrm{c}$. By applying the back-conversion $\mathbf{S}(\mathbf{M}_\mathrm{CPA}^\text{(3)})$ from Eq. (\ref{eq:defSMconversion2}), we can derive the scattering matrix $\mathbf{S}_\mathrm{CPA}^\text{(3)}$ for the entire CPA system. As before, our main focus remains on the top-left quadrant of $\mathbf{S}_\mathrm{CPA}^\text{(3)}$, which represents the CPA's reflection matrix $\mathbf{R}_\mathrm{CPA}^\text{(3)}$ from the left side. 

\noindent After applying the substitutions (\ref{eq:diagsubst}), replacing $\textbf{T}_\text{4f}^2=e^{2i\phi} \mathbb{1}$, and further employing the energy conservation relations $r_1^2 + t_1^2 = 1$, $r_2^2 + t_2^2 = 1$, $r_\mathrm{c}^2 + t_\mathrm{c}^2 = 1$, $t_1^2 = 1 - r_1^2$, and $t_\mathrm{c}^2 = 1 - r_\mathrm{c}^2$, and setting $r_2=1$, the matrix expression for $\mathbf{S}_\mathrm{CPA}^\text{(3)}$ can be reformulated as the following scalar equation: 
\begin{eqnarray}
    r_\mathrm{CPA}^\text{(3)}=
    \frac{r_\mathrm{1}
    + r_\mathrm{c}  \left[r_\mathrm{1} \left(r_\mathrm{c}+t_\mathrm{a}^2\right)+1\right] e^{2 i \phi }    
    + r_\mathrm{c}  \left[t_\mathrm{a}^2 (r_\mathrm{1}+r_\mathrm{c})+1\right] e^{4 i \phi }     
    + t_\mathrm{a}^2 e^{6 i \phi }}
    {1
    + r_\mathrm{c}  \left(r_\mathrm{1}+r_\mathrm{c}+t_\mathrm{a}^2\right) e^{2 i \phi }
    + r_\mathrm{c} \left(r_\mathrm{1} r_\mathrm{c} t_\mathrm{a}^2+r_\mathrm{1}+t_\mathrm{a}^2\right) e^{4 i \phi } 
    +r_\mathrm{1} t_\mathrm{a}^2 e^{6 i \phi }}
\end{eqnarray}

\noindent Similarly to Section S3, we can deduce the conditions for the CPA by solving the following system of equations:
\begin{eqnarray}
\left. \vert r_\mathrm{CPA}^\text{(3)}  \vert^2 \right\vert_{\phi=\frac{\pi}{2}} &= 0 \\
\partial_\phi^2 \left. \vert r_\mathrm{CPA}^\text{(3)}  \vert^2 \right\vert_{\phi=\frac{\pi}{2}} &= 0
\end{eqnarray}
After solving this system of equations, we obtain:
\begin{eqnarray}
&&t_\mathrm{a} = \sqrt{r_1}  \\
&&r_\mathrm{c} =
\frac{\sqrt{1+14 r_1^2 + r_1^4}-r_1^2 - 1}{2 r_1}
\end{eqnarray}
which can be expressed equivalently as
\begin{eqnarray}
&&T_\mathrm{a}  = \sqrt{R_1} \\
&&R_\mathrm{c}= 
\frac{\big(\sqrt{1+14R_1+R_1^2}-R_1-1\big)^2}{4 R_1}
\label{eq:fourteen} 
\end{eqnarray}

\begin{figure}[ht]
\includegraphics[height=4cm]{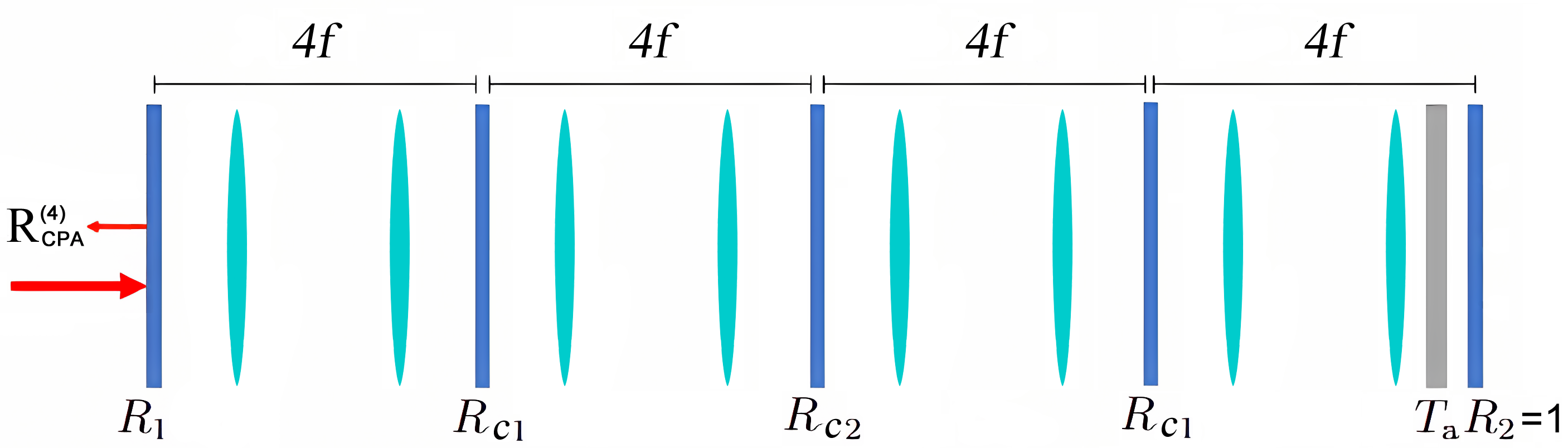}
\caption{\label{fig:EPEPEP-CPA} Creating an MAD-EP-CPA by critically coupling four $4f$-cavities of equal optical length $4f$ each, with the weak absorber situated in the rightmost cavity adjacent to the totally reflective mirror. Note the symmetrical arrangement of the coupling mirrors, where the leftmost and rightmost coupling mirrors possess the same reflectivity $R_\mathrm{c1}$, while the central coupling mirror has a distinct reflectivity $R_\mathrm{c2}$.}
\end{figure}

\noindent Similarly, analytical expressions can be derived to describe the conditions for an EP-CPA consisting of four cavities, as illustrated in Fig.~\ref{fig:EPEPEP-CPA}.

\noindent In this case, since the leftmost and rightmost coupling mirrors share the same reflectivity, denoted as $R_\mathrm{c1}$, while the central coupling mirror has a different reflectivity  $R_\mathrm{c2}$, we must define two distinct scattering matrices $\mathbf{S}_\mathrm{c1}$ and $\mathbf{S}_\mathrm{c2}$, from which we can derive the two corresponding transfer matrices $\mathbf{M}_\mathrm{c1}$ and $\mathbf{M}_\mathrm{c2}$. These will be used to compute the total transfer matrix of the system.
\begin{equation}
    \mathbf{M}_\mathrm{CPA}^{(4)} = \mathbf{M}_1 \mathbf{M}_\mathrm{4f} 
    \mathbf{M}_\mathrm{c1} \mathbf{M}_\mathrm{4f} \mathbf{M}_\mathrm{c2} \mathbf{M}_\mathrm{4f} \mathbf{M}_\mathrm{c1} \mathbf{M}_\mathrm{4f}
    \mathbf{M}_\mathrm{a} \mathbf{M}_2
\end{equation}
The scattering matrix $\mathbf{S}_\mathrm{CPA}^{(4)}$ for the entire CPA system can be obtained by using the back-conversion $\mathbf{S}(\mathbf{M}^{(4)})$ from Eq. (\ref{eq:defSMconversion2}). This matrix contains the CPA’s reflection matrix $\mathbf{R}_\mathrm{CPA}^{(4)}$ in the top-left quadrant. After making analogous substitutions as previously, we arrive at:

\begin{tiny}
\begin{eqnarray}
    r_\text{CPA}^{(4)} =  
    \frac{r_\mathrm{1}
    +r_\mathrm{c1}  \left[r_\mathrm{1} \left(2 r_\mathrm{c2}+t_\mathrm{a}^2\right)+1\right]e^{2 i \phi }    
    + \left[t_\mathrm{a}^2 \left(r_\mathrm{1} \left(r_\mathrm{c1}^2+1\right) r_\mathrm{c2}+r_\mathrm{c1}^2\right)+r_\mathrm{c1}^2 (r_\mathrm{1}+r_\mathrm{c2})+r_\mathrm{c2}\right] e^{4 i \phi } 
    +r_\mathrm{c1}  \left[t_\mathrm{a}^2 (r_\mathrm{1}+2 r_\mathrm{c2})+1\right] e^{6 i \phi }    
    +t_\mathrm{a}^2 e^{8 i \phi }}
    {1 
    +r_\mathrm{c1}  \left(r_\mathrm{1}+2 r_\mathrm{c2}+t_\mathrm{a}^2\right) e^{2 i \phi }
    + \left[r_\mathrm{c1}^2 \left(t_\mathrm{a}^2 (r_\mathrm{1}+r_\mathrm{c2})+r_\mathrm{1} r_\mathrm{c2}+1\right)+r_\mathrm{c2} \left(r_\mathrm{1}+t_\mathrm{a}^2\right)\right] e^{4 i \phi }    
    +r_\mathrm{c1}  \left(2 r_\mathrm{1} r_\mathrm{c2} t_\mathrm{a}^2+r_\mathrm{1}+t_\mathrm{a}^2\right) e^{6 i \phi }
    +r_\mathrm{1} t_\mathrm{a}^2 e^{8 i \phi }}
    \;\;\;\;
\end{eqnarray}
\end{tiny}

\noindent With the addition of another cavity, the system of equations necessary to determine the CPA conditions now incorporates an additional equation:
\begin{eqnarray}
\left. \vert r_\mathrm{CPA}^{(4)}\vert^2 \right\vert_{\phi=\frac{\pi}{2}} &= 0 \\
\partial_\phi^2 \left. \vert r_\mathrm{CPA}^{(4)}\vert^2 \right\vert_{\phi=\frac{\pi}{2}} &= 0 \\
\partial_\phi^4 \left. \vert r_\mathrm{CPA}^{(4)}\vert^2 \right\vert_{\phi=\frac{\pi}{2}} &= 0
\end{eqnarray}

\noindent While it is possible to express the exact solution analytically, doing so requires some effort:
\begin{eqnarray}
&t_\mathrm{a} = \sqrt{r_1} \\
&r_\mathrm{c1} =
    \frac{D}{2^{1/3} A}-
    \frac{2^{1/3} B}{A D}+\frac{4 \left(r_\mathrm{1}^5-r_\mathrm{1}\right)}{A} 
\end{eqnarray}
with
\begin{footnotesize}
\begin{eqnarray}
A &=& 3 \left(r_\mathrm{1}^6-3 r_\mathrm{1}^4+3 r_\mathrm{1}^2-1  \right) \nonumber  \\
B &=& 3 r_\mathrm{1}^{12}+14 r_\mathrm{1}^{10}-147 r_\mathrm{1}^8+260 r_\mathrm{1}^6-147 r_\mathrm{1}^4+14 r_\mathrm{1}^2+3 \nonumber \\
C &=& 72 r_\mathrm{1}^{17}-880 r_\mathrm{1}^{15}+3312 r_\mathrm{1}^{13}-4272 r_\mathrm{1}^{11}+4272 r_\mathrm{1}^7-3312 r_\mathrm{1}^5+880 \nonumber 
r_\mathrm{1}^3-72 r_\mathrm{1} \\
D &=& \left(C+\sqrt{4B^3+C^2}\right)^{1/3}\nonumber 
\end{eqnarray}
\end{footnotesize}

\noindent And, finally:
\begin{eqnarray}
&r_\mathrm{c2} =
\frac{
D \left(C^5 I + J K\right)
+8 C^6 F r_\mathrm{1}  \left(r_\mathrm{1}^2+1\right)
-2 D^{2} H \left[3 J+4 C^3 G r_\mathrm{1} \left(r_\mathrm{1}^2+1\right) \right]
-\sqrt{3} \sqrt{B} C^7 E
}
{6 D^{5} r_\mathrm{1} \left(r_\mathrm{1}^4-1\right)}
\end{eqnarray}
with
\begin{footnotesize}
\begin{eqnarray}
A &=& 36 r_1^{17}-440 r_1^{15}+1656 r_1^{13}-2136 r_1^{11}+2136 r_1^7-1656 r_1^5+440 r_1^3-36 r_1 \nonumber \\
B &=& r_1^8+76 r_1^6-282 r_1^4+76 r_1^2+1 \nonumber \\
C &=& r_1^2-1 \nonumber \\
D &=& \left(A-3 \sqrt{3} \sqrt{B} C^7\right)^{1/3} \nonumber \\
E &=& \left( 9 r_1^{16}-168 r_1^{14}+2108 r_1^{12}-6040 r_1^{10}+4086 r_1^8-6040 r_1^6+2108 r_1^4-168 r_1^2+9\right) (r_\mathrm{1}-1) (r_\mathrm{1}+1) \nonumber  \\
F &=& 9 r_1^{20}-462 r_1^{18}+5045 r_1^{16}-32232 r_1^{14}+115266 r_1^{12}-191636 r_1^{10}+115266 r_1^8-32232 r_1^6+5045 r_1^4-462 r_1^2+9 \nonumber \\
G &=& 9 r_1^8-92 r_1^6+230 r_1^4-92 r_1^2+9 \nonumber  \\
H &=& r_1^6-5 r_1^4+5 r_1^2-1 \nonumber \\
I &=& 9 r_1^{20}+1206 r_1^{18}-13547 r_1^{16}+49032 r_1^{14}-56862 r_1^{12}+7556 r_1^{10}-56862 r_1^8+49032 r_1^6-13547 r_1^4+1206 r_1^2+9 \nonumber \\
J &=& \sqrt{3} \sqrt{B C^{14}} \nonumber \\
K &=&   84 r_\mathrm{1}^{11} -572  r_\mathrm{1}^9 +744  r_\mathrm{1}^7 +744  r_\mathrm{1}^5 -572 r_\mathrm{1}^3 +84  r_\mathrm{1}   \nonumber 
\end{eqnarray}
\end{footnotesize}

\noindent By performing a series expansion around $r_1=1$, we can approximate the expression for $r_\mathrm{c1}$ with the following simplified formula, which maintains good fidelity for all values\linebreak $0.5 < r_1 < 1$:
\begin{eqnarray}
r_\mathrm{c1} \approx 1 + \frac{1-\sqrt{2}}{2}\left( 2 - 5 r_1 + 4 r_1^2 - r_1^3\right) + \frac{5-4\sqrt{2}}{2} (r_1 - 1)^4 -\frac{4-3\sqrt{2}}{2} (r_1 - 1)^5
\end{eqnarray}

\noindent Furthermore, by employing numerical parameter fitting of a fit-function to the exact solution, we can derive an approximate formula for $r_\mathrm{c2}$ that is suitable for all values $0.5 < r_1 < 1$:
\begin{eqnarray}
r_\mathrm{c2} \approx 1 - 0.0984341 (1 - r_1)^2 + 0.0197372 (1 - r_1)^3 - 0.327913 (1 - r_1)^4 
\end{eqnarray}
For an EP-CPA that incorporates five or more cavities, it is no longer possible to represent the exact solutions analytically. 

\begin{figure}[h]
\includegraphics[width=12cm]{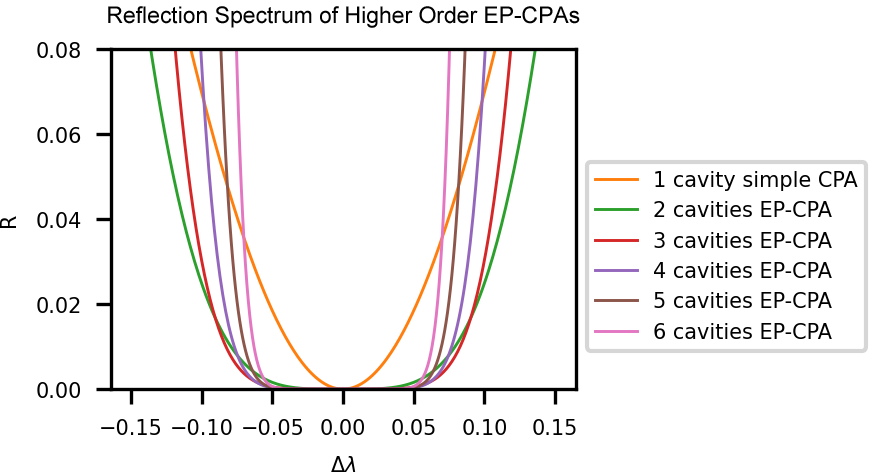}
\caption{\label{fig:MCCPAplot} Comparison of the reflection spectrum of a Conventional CPA with various EP-CPAs consisting of multiple cavities. The power reflectivity of the input coupling mirror is $R_1=0.7$. }
\end{figure}

\noindent When generalizing the concept illustrated in Fig.~\ref{fig:EPEPEP-CPA} by coupling $n>4$ cavities, $m=n-1$ coupling mirrors with reflection coefficients $R_\mathrm{c1}, R_\mathrm{c2}, ..., R_\mathrm{cm}$ are required. Importantly, for spectral broadening to occur, symmetrical pairs of internal coupling mirrors must have identical reflection coefficients, such that $R_\mathrm{c1} = R_\mathrm{cm}, R_\mathrm{c2} = R_\mathrm{c(m-1)}, $, and so on. However, as the number of coupled cavities in an EP-CPA increases, the reflection spectrum around the resonance frequencies starts to resemble a Butterworth bandpass filter, as known from electrical filter theory. This observation can be supported by plotting the reflectance spectra $R(\Delta \lambda)$ in Fig.~\ref{fig:MCCPAplot} as a function of a phase $\phi$ by substituting $\Delta \lambda \rightarrow 2 \pi L / \phi$, and subsequently comparing $R(\phi)$ with $R = 1-\mathrm{b}(\phi)$, where  
\begin{eqnarray}
&&\mathrm{b}(\phi)  = \frac{1}{1+\left(\frac{\phi}{\phi_{NN}}\right)^{2n}} \label{eq:but} 
\end{eqnarray}
denotes the conventional Butterworth equation, as recognized in electrical theory, $\phi_{NN}$ is the half-maximum-reflection phase-shift of the corresponding WLC-CPA, and $n$ is the number of coupled cavities. This is consistent with the known fact, as mentioned in [4], that the Butterworth-function can also describe the transmission behavior of a multimirror Fabry-Perot interferometer without a dissipating element, which is a configuration related to the EP-CPAs under discussion. The crucial takeaway from this analysis is that, unless the objective is to construct a Butterworth-like band-absorption-filter, augmenting the number of cavities beyond three does not substantially enhance the broadband absorption behavior 

\section{S6. Deriving the CPA conditions for the alternative EP-CPA design and Quartic Behavior Proof}

\noindent In this section, we will outline the procedure to derive CPA conditions for the alternative EP-CPA Design~B illustrated in Fig.~3(a) of the main text and prove the quartic beavior. The calculation will be based on the beam-path depicted in this figure. In our previous calculations, to determine the effect of an optical element (e.g., a mirror), it was sufficient to consider the amplitude vectors of the left-incident light field $\mathbf{a}^{\mathrm{left}}_{\mathrm{in}}$ and the right-incident light field $\mathbf{a}^{\mathrm{right}}_{\mathrm{in}}$. We then calculated the corresponding outgoing amplitude vectors $\mathbf{a}^{\mathrm{left}}_{\mathrm{out}}$ and $\mathbf{a}^{\mathrm{right}}_{\mathrm{out}}$ using a $4 \times 4$ scattering block-matrix, as shown in Eq.~(\ref{eq:smatrix}). However, the scenario  becomes more intricate for the case at hand. It necessitates considering for each optical element \textit{two} distinct fields of light incident from the left and right, and calculating \textit{two} separate outgoing fields for the left and right (i.e., their corresponding amplitude vectors). This is accomplished by expanding the amplitude vectors $\mathbf{a}^{\mathrm{left}}_{\mathrm{in}}$, $\mathbf{a}^{\mathrm{right}}_{\mathrm{in}}$, $\mathbf{a}^{\mathrm{left}}_{\mathrm{out}}$, and $\mathbf{a}^{\mathrm{right}}_{\mathrm{out}}$ to to have two components each, where each component is itself an amplitude vector. These components  will represent the amplitudes of the light field above the optical axis (first vector element with index $1$), and the amplitudes of the light field below the optical axis (second vector element with index $2$):
\begin{equation}
    \mathbf{a}^{\mathrm{left}}_{\mathrm{in}} = \begin{pmatrix} {\mathbf{a}_1}^{\mathrm{left}}_{\mathrm{in}} \\ {\mathbf{a}_2}^{\mathrm{left}}_{\mathrm{in}} \end{pmatrix} \;\;\;
    \mathbf{a}^{\mathrm{right}}_{\mathrm{in}} = \begin{pmatrix} {\mathbf{a}_1}^{\mathrm{right}}_{\mathrm{in}} \\ {\mathbf{a}_2}^{\mathrm{right}}_{\mathrm{in}}    
    \end{pmatrix} \;\;\;
     \mathbf{a}^{\mathrm{left}}_{\mathrm{out}} = \begin{pmatrix} {\mathbf{a}_1}^{\mathrm{left}}_{\mathrm{out}} \\ {\mathbf{a}_2}^{\mathrm{left}}_{\mathrm{out}} \end{pmatrix} \;\;\;
    \mathbf{a}^{\mathrm{right}}_{\mathrm{out}} = \begin{pmatrix} {\mathbf{a}_1}^{\mathrm{right}}_{\mathrm{out}} \\ {\mathbf{a}_2}^{\mathrm{right}}_{\mathrm{out}}    
    \end{pmatrix}
    \label{eq:aVec}
\end{equation}
Consequently, the scattering matrices we will be working with have evolved into $2\times 2$ block matrices, where each element is itself a $2\times 2$ block matrix, transforming them into $4\times 4$  block matrices:
\begin{equation}
    \left( \begin{array}{ll} \mathbf{a}^{\mathrm{left}}_{\mathrm{out}} \\ 
    \mathbf{a}^{\mathrm{right}}_{\mathrm{out}}  \end{array} \right) = 
    \mathbf{S} 
    \left( \begin{array}{ll} \mathbf{a}^{\mathrm{left}}_{\mathrm{in}} \\ 
    \mathbf{a}^{\mathrm{right}}_{\mathrm{in}}  \end{array} \right) = 
    \left( \begin{array}{cc} \mathbf{R}_\mathrm{L} & \mathbf{T}_\mathrm{RL} \\ \mathbf{T}_\mathrm{LR} & \mathbf{R}_\mathrm{R} \end{array} \right)
    \left( \begin{array}{ll} \mathbf{a}^{\mathrm{left}}_{\mathrm{in}} \\ 
    \mathbf{a}^{\mathrm{right}}_{\mathrm{in}}  \end{array} \right)= 
    \left( \begin{array}{cccc}
    \mathbf{R}_\mathrm{L}^{11} & \mathbf{R}_\mathrm{L}^{12} & \mathbf{T}_\mathrm{RL}^{11} & \mathbf{T}_\mathrm{RL}^{12} \\
    \mathbf{R}_\mathrm{L}^{21} & \mathbf{R}_\mathrm{L}^{22} & \mathbf{T}_\mathrm{RL}^{21} & \mathbf{T}_\mathrm{RL}^{22} \\
    \mathbf{T}_\mathrm{LR}^{11} & \mathbf{T}_\mathrm{LR}^{12} & \mathbf{R}_\mathrm{R}^{11} & \mathbf{R}_\mathrm{R}^{12} \\
   \mathbf{T}_\mathrm{LR}^{21} & \mathbf{T}_\mathrm{LR}^{22} & \mathbf{R}_\mathrm{R}^{21} & \mathbf{R}_\mathrm{R}^{22} 
    \end{array} \right)
    \left( \begin{array}{ll} \mathbf{a}^{\mathrm{left}}_{\mathrm{in}} \\ 
    \mathbf{a}^{\mathrm{right}}_{\mathrm{in}}  \end{array} \right)
    \label{eq:smatrix2}
\end{equation}
Although we aim to maintain the convention established in Eq.~(\ref{eq:defS2}), it is now necessary to reformulate it for the $4 \times 4$ scattering block matrix:
\begin{eqnarray}
\mathbf{S}=\begin{pmatrix}
\mathbf{R}_\mathrm{L} & \mathbf{T}_\mathrm{RL} \\
 \mathbf{T}_\mathrm{LR} & \mathbf{R}_\mathrm{R}
\end{pmatrix} =
\begin{pmatrix}
\mathbf{R} & \mathbf{T} \\ \mathbf{T} & -\mathbf{R} 
\end{pmatrix} =
\begin{pmatrix}
\mathbf{R}_{11} & \mathbf{R}_{12} & \mathbf{T}_{11} & \mathbf{T}_{12} \\
\mathbf{R}_{21} & \mathbf{R}_{22} & \mathbf{T}_{21} & \mathbf{T}_{22} \\
\mathbf{T}_{11} & \mathbf{T}_{12} & -\mathbf{R}_{11} & -\mathbf{R}_{12} \\
\mathbf{T}_{21} & \mathbf{T}_{22} & -\mathbf{R}_{21} & -\mathbf{R}_{22} 
\end{pmatrix}
\;\;\;  \mathbf{R}_{ij}, \mathbf{T}_{ij} \in \mathbb{R}^{n \times n} \label{eq:defS3}
\end{eqnarray}
Given these modifications, we are now in a position to write down the $4 \times 4$ scattering block matrix $\mathbf{S}_\mathrm{ML}$ for the left ``split'' mirror:
\begin{eqnarray}
\mathbf{S}_\mathrm{ML}=
\begin{pmatrix}
\mathbf{R}_\mathrm{ML1} & \mathbb{0} & \mathbf{T}_\mathrm{ML1} & \mathbb{0} \\
\mathbb{0} & \mathbf{R}_\mathrm{ML2} & \mathbb{0} & \mathbf{T}_\mathrm{ML2} \\
\mathbf{T}_\mathrm{ML1} & \mathbb{0} & -\mathbf{R}_\mathrm{ML1} & \mathbb{0} \\
\mathbb{0} & \mathbf{T}_\mathrm{ML2} & \mathbb{0} & -\mathbf{R}_\mathrm{ML2} 
\end{pmatrix} \label{eq:defSML}
\end{eqnarray}
It is important to note that $\mathbf{S}_\mathrm{ML}$ includes separate reflection- and transmission matrices for the ``upper'' and the ``lower'' light fields. In contrast, we do not need to distinguish between these two spatially separated light fields in the scattering matrix for the central mirror $\mathbf{S}_\mathrm{MC}$ and the right mirror $\mathbf{S}_\mathrm{MR}$, since each of these mirrors 
 reflects and transmits in the same way above and below the optical axis:
\begin{eqnarray}
\mathbf{S}_\mathrm{MC}=
\begin{pmatrix}
\mathbf{R}_\mathrm{MC} & \mathbb{0} & \mathbf{T}_\mathrm{MC} & \mathbb{0} \\
\mathbb{0} & \mathbf{R}_\mathrm{MC} & \mathbb{0} & \mathbf{T}_\mathrm{MC} \\
\mathbf{T}_\mathrm{MC} & \mathbb{0} & -\mathbf{R}_\mathrm{MC} & \mathbb{0} \\
\mathbb{0} & \mathbf{T}_\mathrm{MC} & \mathbb{0} & -\mathbf{R}_\mathrm{MC} 
\end{pmatrix}
\;\;\;\;\;
\mathbf{S}_\mathrm{MR}=
\begin{pmatrix}
\mathbf{R}_\mathrm{MR} & \mathbb{0} & \mathbf{T}_\mathrm{MR} & \mathbb{0} \\
\mathbb{0} & \mathbf{R}_\mathrm{MR} & \mathbb{0} & \mathbf{T}_\mathrm{MR} \\
\mathbf{T}_\mathrm{MR} & \mathbb{0} & -\mathbf{R}_\mathrm{MR} & \mathbb{0} \\
\mathbb{0} & \mathbf{T}_\mathrm{MR} & \mathbb{0} & -\mathbf{R}_\mathrm{MR} 
\end{pmatrix} \label{eq:defSC_SMR}
\end{eqnarray}
Given that the absorber only affects the amplitude of the light field below the optical axis, characterized by a transmission matrix $\mathbf{T}_\mathrm{a}$, we can express the corresponding $4 \times 4$ scattering block matrix as follows:
\begin{eqnarray}
\mathbf{S}_\mathrm{a}=
\begin{pmatrix}
\mathbb{0} & \mathbb{0} & \mathbb{1} & \mathbb{0} \\
\mathbb{0} & \mathbb{0} & \mathbb{0} & \mathbf{T}_\mathrm{a} \\
\mathbb{1} & \mathbb{0} & \mathbb{0} & \mathbb{0} \\
\mathbb{0} & \mathbf{T}_\mathrm{a} & \mathbb{0} & \mathbb{0} 
\end{pmatrix} \label{eq:defSa}
\end{eqnarray}
 Let $\textbf{T}_\text{2f}$ represent the transmission matrix for propagation between the focal points of a single lens (hence, representing a Fourier transformation which also applies a uniform phase $e^{i \phi}$ to all modes). Then, the free-space propagation between each optical element, spanning a distance of one focal length $f$, can be represented by the following  $4 \times 4$ scattering block matrix.
\begin{eqnarray}
\mathbf{S}_\mathrm{2f}=
\begin{pmatrix}
\mathbb{0} & \mathbb{0} & \mathbf{T}_\mathrm{2f} & \mathbb{0} \\
\mathbb{0} & \mathbb{0} & \mathbb{0} & \mathbf{T}_\mathrm{2f} \\
\mathbf{T}_\mathrm{2f} & \mathbb{0} & \mathbb{0} & \mathbb{0} \\
\mathbb{0} & \mathbf{T}_\mathrm{2f} & \mathbb{0} & \mathbb{0} 
\end{pmatrix} \label{eq:defS2f}
\end{eqnarray}
The $4 \times 4$ scattering block matrix $\mathbf{S}_\mathrm{LL}$ representing the left lens must represent the behavior demonstrated by the exemplary beam path in  Fig.~4(a) of the main paper, indicating that the left incident light field shifts from above to below the optical axis upon reaching the central mirror, and vice versa. Consequently, the lower-left $2 \times 2$ sub-block-matrix, which represents the left-to-right-transmission, is a block matrix with identity matrices in the counter-diagonal elements. In contrast, the right-to-left propagating light field does not change sides with respect to the optical axis. Therefore, the top-right $2 \times 2$ sub-block-matrix, representing the right-to-left transmission, is a block matrix with identity matrices in the main-diagonal elements, making it an identity matrix:
\begin{eqnarray}
\mathbf{S}_\mathrm{LL}=
\begin{pmatrix}
\mathbb{0} & \mathbb{0} & \mathbb{1} & \mathbb{0} \\
\mathbb{0} & \mathbb{0} & \mathbb{0} & \mathbb{1} \\
\mathbb{0} & \mathbb{1} & \mathbb{0} & \mathbb{0} \\
\mathbb{1} & \mathbb{0} & \mathbb{0} & \mathbb{0} 
\end{pmatrix} \label{eq:defSLL}
\end{eqnarray}
In the case of the right lens, the situation is reversed: the left-to-right propagating light field does not change sides, while the right-to-left propagating light field shifts from above to below the optical axis upon reaching the central mirror (and vice versa). As a result, the $4 \times 4$ scattering block matrix $\mathbf{S}_\mathrm{LR}$ for the right lens is structured accordingly:
\begin{eqnarray}
\mathbf{S}_\mathrm{LR}=
\begin{pmatrix}
\mathbb{0} & \mathbb{0} & \mathbb{0} & \mathbb{1} \\
\mathbb{0} & \mathbb{0} & \mathbb{1} & \mathbb{0} \\
\mathbb{1} & \mathbb{0} & \mathbb{0} & \mathbb{0} \\
\mathbb{0} & \mathbb{1} & \mathbb{0} & \mathbb{0} 
\end{pmatrix} \label{eq:defSLR}
\end{eqnarray}
To convert the previously defined scattering matrices into their corresponding transfer matrices with the goal to compute the total transfer matrix $\mathbf{M}_\mathrm{CPA}$ for the entire CPA, the conversion function $\mathbf{M}(\mathbf{S})$, as defined in  (\ref{eq:defSMconversion2}), can be utilized once more. This time, however, the multiplication rule for $2\times 2$ block-matrices, as detailed in (\ref{eq:BMmult2}), is required to apply the conversion functions $\mathbf{M}(\mathbf{S})$ and $\mathbf{S}(\mathbf{M})$ from Eqs. (\ref{eq:defSMconversion2}). Additionally, calculating the inverse of $2 \times 2$ block matrices becomes essential. Following [5,~p.~184] and [6], we do this as follows: Let $\mathbf{M}$ be a $2 \times 2$ block matrix with the components $\mathbf{M}_{11}=\mathbf{A}$, $\mathbf{M}_{12}=\mathbf{B}$, $\mathbf{M}_{21}=\mathbf{C}$, and $\mathbf{M}_{22}=\mathbf{D}$, each being matrices themselves. Then 

\begin{align}
    \mathbf{M}^{-1}=
    \begin{cases}
    \begin{pmatrix}
    \mathbb{0} & \mathbf{C}^{-1} \\
     \mathbf{B}^{-1} & \mathbb{0}
    \end{pmatrix}, &\text{for} \; \mathbf{A}=\mathbb{0} \wedge  \mathbf{D}=\mathbb{0} \\
    \noalign{\vskip9pt}
    \begin{pmatrix}
    \mathbf{A}^{-1} & \mathbb{0} \\
    \mathbb{0} &  \mathbf{D}^{-1} 
    \end{pmatrix}, &\text{for} \; \mathbf{B}=\mathbb{0} \wedge  \mathbf{C}=\mathbb{0} \\
    \noalign{\vskip9pt}
    \begin{pmatrix}
    (\mathbf{A}-\mathbf{B} \mathbf{D}^{-1} \mathbf{C})^{-1} &
    -\mathbf{A}^{-1} \mathbf{B} ( \mathbf{D} - \mathbf{C} \mathbf{A}^{-1} \mathbf{B})^{-1} \\
    -\mathbf{D}^{-1} \mathbf{C}
    (\mathbf{A}-\mathbf{B}\mathbf{D}^{-1} \mathbf{C})^{-1}&
    (\mathbf{D} - \mathbf{C} \mathbf{A}^{-1}
    \mathbf{B})^{-1}
    \end{pmatrix}, &\text{for} \;
    \mathbf{A}, \mathbf{D} \; \text{nonsingular}
    \end{cases}
    \label{eq:block_matrix_inverse}
\end{align}

\noindent After converting all $4 \times 4$ scattering block matrices into the corresponding $4 \times 4$ transfer block matrices using the $\mathbf{M}(\mathbf{S})$ function, we can succinctly express the transfer matrix $\mathbf{M}_\mathrm{CPA}^\text{(B)}$ for the entire CPA (Design~B):
\begin{eqnarray}
\mathbf{M}_\mathrm{CPA}^\text{(B)}=
\mathbf{M}_\mathrm{ML} \mathbf{M}_\mathrm{a} \mathbf{M}_\mathrm{LL} 
\mathbf{M}_\mathrm{2f} \mathbf{M}_\mathrm{MC}  \mathbf{M}_\mathrm{LR}  \mathbf{M}_\mathrm{2f} \mathbf{M}_\mathrm{MR}
\label{eq:MCPA2}
\end{eqnarray}
Given that all the $\mathbf{M}$-matrices in Eq.~(\ref{eq:MCPA2}) are now $4 \times 4$ block matrices, where each element is a distinct matrix in its own right, the following multiplication rule for $4 \times 4$ block matrices, $\mathbf{X}$ and $\mathbf{Y}$, must be applied during each multiplication for every matrix element in the calculation of expression (\ref{eq:MCPA2}):
\begin{equation}
    \left( \mathbf{X} \mathbf{Y}\right)_{ij} = \sum_{k=1}^4{\mathbf{X}_{ik} \mathbf{Y}_{kj}}
    \label{eq:BMmult4}
\end{equation}
The corresponding scattering matrix can then be obtained from $\mathbf{M}_\mathrm{CPA}^\text{(B)}$ by utilizing the $\mathbf{S}(\mathbf{M})$ back-conversion function, again taking into account the $2 \times 2$ block matrix multiplication rule (\ref{eq:BMmult2}) and the the $2 \times 2$ block matrix inversion method (\ref{eq:block_matrix_inverse}) when calculating
\begin{eqnarray}
\mathbf{S}_\mathrm{CPA}^\text{(B)} = \mathbf{S}(\mathbf{M}_\mathrm{CPA}^\text{(B)}).
\label{eq:SCPA2}
\end{eqnarray}
Our interest lies in examining the CPA's behavior with respect to reflections on its left side. Hence, we focus on the top-left quadrant $2 \times 2$ block matrix of  $\mathbf{S}_\mathrm{CPA}^\text{(B)}$. In particular, we are concerned with the CPA’s reflection behavior for \textit{incident light above the optical axis}. Consequently, we must extract the top-left entry of this sub-matrix, which also coincides with the top-left entry of the entire scattering matrix $\mathbf{S}_\mathrm{CPA}^\text{(B)}$. This entry contains the reflection matrix $\mathbf{R}_\mathrm{CPA}^\text{(B)}$ that we are interested in. Assuming again that the same reflection and transmission coefficients apply to all modes at the mirrors and the absorber, then all matrices except $\textbf{T}_\text{f}$ become diagonal matrices:
\begin{equation}
\begin{aligned} 
    \textbf{R}_\text{ML1} = r_{1a} \mathbb{1}, \;\;
    \textbf{R}_\text{ML2} = r_{1b} \mathbb{1}, \;\;
    &\textbf{T}_\text{ML1} = t_{1a} \mathbb{1}, \;\;
    \textbf{T}_\text{ML2} = r_{1b} \mathbb{1} \\
    \textbf{R}_\text{MC} = r_\text{c} \mathbb{1}, \;\;
    \textbf{T}_\text{MC} = t_\text{c} \mathbb{1}, \;\;
    \textbf{R}_\text{MR} &= r_2 \mathbb{1}, \;\;
    \textbf{T}_\text{MR} = t_2 \mathbb{1}, \;\;
    \textbf{T}_\text{a} = t_\text{a} \mathbb{1}
    \end{aligned}
    \label{eq:diagsubst2}
\end{equation}

\noindent After implementing the aforementioned substitutions, and further applying the energy conservation substitutions $r_\mathrm{1a}^2+t_\mathrm{1a}^2 = 1$, $r_\mathrm{1b}^2+t_\mathrm{1b}^2 = 1$, $r_\mathrm{c}^2+t_\mathrm{c}^2 = 1$, and $r_2^2+t_2^2 = 1$, then setting $r_\text{1b}=1$ and $r_2=1$, and finally substituting $t_\text{c}^2 = 1 - r_\text{c}^2$ and $t_\text{1a}^2 = 1 - r_\text{1a}^2$, a  simplified matrix equation for $\textbf{R}_\text{CPA}^\text{(B)}$ can be derived:
\begin{align}
    \textbf{R}_\text{CPA}^\text{(B)} = \frac{t_\text{a}^2 \, \textbf{T}_\text{2f}^{8} + [ 1 + r_\text{1a} t_\text{a}^2  - r_\text{c}^2(t_\text{a}^2+1)(r_\text{1a}^2+1) ] \, \textbf{T}_\text{2f}^4  + r_\text{1a} \mathbb{1} }{r_\text{1a} t_\text{a}^2 \, \textbf{T}_\text{2f}^{8} + [r_\text{1a} + t_\text{a}^2  - r_\text{c}^2(t_\text{a}^2+1)(r_\text{1a}^2+1) ] \, \textbf{T}_\text{2f}^{4} + \mathbb{1}}
    \label{eq:RLa}
\end{align}
This correponds to Eq.~(3) in the main paper. Observe how the structure of Eq.~(\ref{eq:RLa}) resembles the structure of Eq.~(\ref{eq:RCPA_simplified2}). Because $\textbf{T}_\text{2f}^4 = \textbf{T}_\text{4f}^2$, the matrix $\textbf{T}_\text{2f}^4$ in Eq. (\ref{eq:RLa}) results in a quadruple Fourier transformation with a uniform phase shift $e^{2i\phi}$ for all modes, and can thus be represented as a simple diagonal matrix $\textbf{T}_\text{4f}^2=e^{2i\phi} \mathbb{1}$. Analogously, $\textbf{T}_\text{2f}^{8}=e^{4i\phi} \mathbb{1}$. Consequently, $ \textbf{R}_\text{CPA}^\text{(B)}$ becomes a diagonal matrix $ \textbf{R}_\text{CPA}^\text{(B)} = r_\text{CPA}^\text{(B)} \mathbb{1}$, and can therefore be reduced to a simple scalar equation, which proves the spatial degeneracy of Design~B.
\begin{eqnarray}
r_\mathrm{CPA}^\text{(B)}=
\frac{t_\text{a}^2 \, e^{4 i \phi} + [ 1 + r_\text{1a} t_\text{a}^2  - r_\text{c}^2(t_\text{a}^2+1)(r_\text{1a}^2+1) ] \, e^{2 i \phi}  + r_\text{1a} }{r_\text{1a} t_\text{a}^2 \, e^{4 i \phi} + [r_\text{1a} + t_\text{a}^2  - r_\text{c}^2(t_\text{a}^2+1)(r_\text{1a}^2+1) ] \, e^{2 i \phi} + 1} \label{eq:rcpaB}
\end{eqnarray}

\begin{figure}[!ht]
\includegraphics[width=12cm]{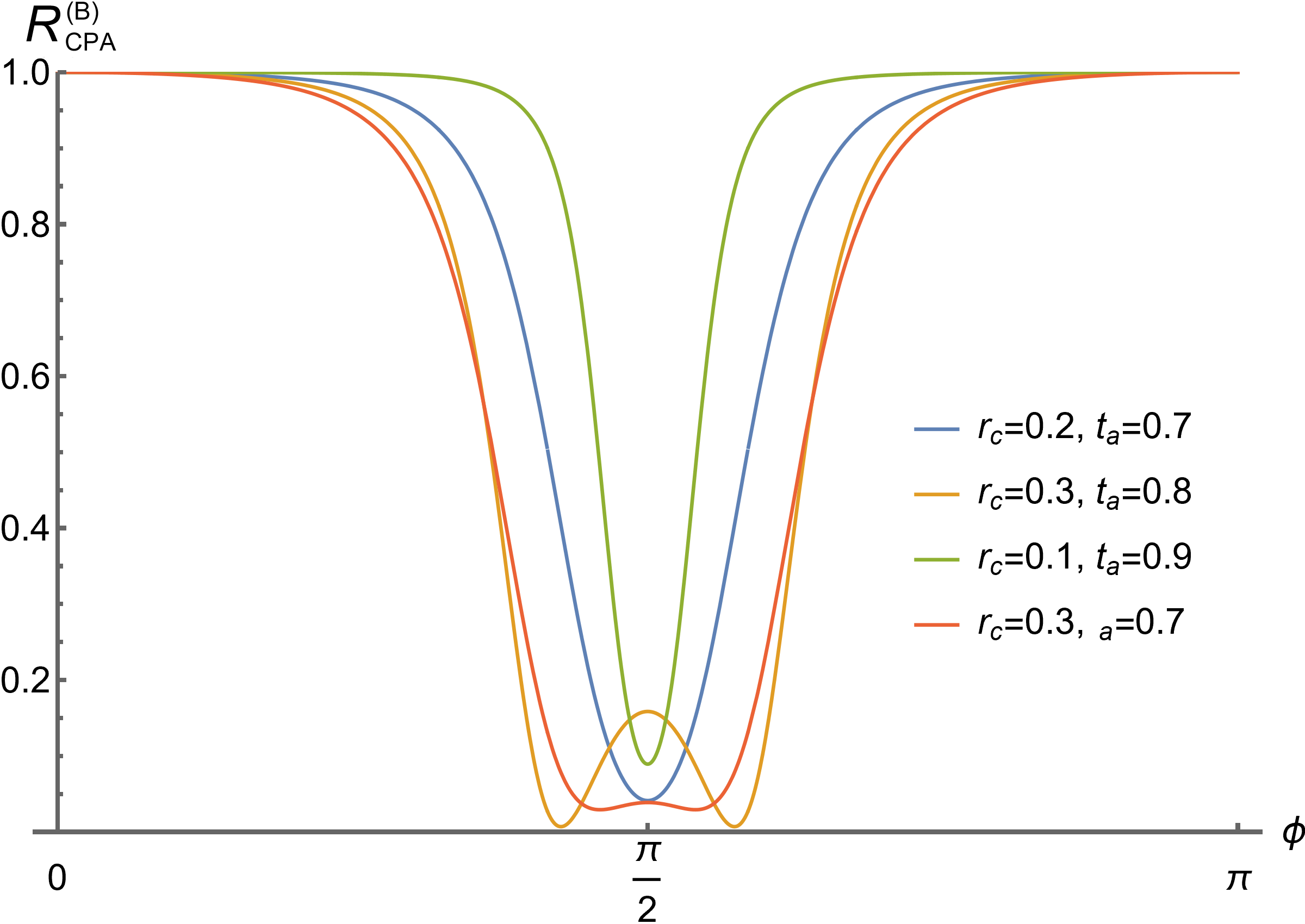}
\caption{\label{fig:deriv_48_CPA} A plot of $R_\mathrm{CPA}^\text{(B)}$ with $r_1=0.7$ and some arbitary values for $r_\mathrm{c}$ and $ t_\mathrm{a}$ shows that we can expect to find perfect absorption at $\phi=\frac{\pi}{2}$, provided we find the right values for $r_\text{c}$ and $t_\text{a}$.} 
\end{figure}

\noindent Plotting $R_\mathrm{CPA}^\text{(B)}(\phi)= \vert r_\mathrm{CPA}^\text{(B)} \vert^2$ (Fig.~\ref{fig:deriv_48_CPA}), we observe that perfect absorption at $\phi=\frac{\pi}{2}$ seems to be possible, contingent on our ability to identify the correct parameter values.
 Following the same reasoning as in section S3, we establish the subsequent system of equations:
\begin{align}
\left. R_\mathrm{CPA}^\text{(B)}\left(\phi\right)\right\vert_{\phi=\frac{\pi}{2}} &= 0 \\
\partial_\phi^2 \left. R_\mathrm{CPA}^\text{(B)}\left(\phi\right)\right\vert_{\phi=\frac{\pi}{2}} &= 0
\end{align}
Solving this set of equations yields the following results:
\begin{align}
t_\mathrm{a} &=\sqrt{r_\mathrm{ML1}} \label{eq:ta48}\\
r_\mathrm{c} &= \frac{1-r_\mathrm{ML1}}{1+r_\mathrm{ML1}} \label{eq:rc48}
\end{align}
Squaring both sides of Eqs. (\ref{eq:ta48}) and (\ref{eq:rc48}), and substituting $t_\mathrm{a}^2 = T_\mathrm{a}$, $r_\mathrm{c}^2 = R_\mathrm{c}$, $r_\mathrm{ML1} = \sqrt{R_\mathrm{1a}}$ results in the CPA conditions for the alternative design as presented in the main paper:
\begin{align}
T_\mathrm{a} = \sqrt{R_\mathrm{1a}} \label{eq:Ta48} 
\;\; &\Leftrightarrow \;\;
R_\mathrm{1a} = T_\mathrm{a}^2
\\
R_\mathrm{c} = \left( \frac{1-\sqrt{R_\mathrm{1a}}}{1+\sqrt{R_\mathrm{1a}}} \right)^2 
\;\; &\Leftrightarrow \;\;
R_\mathrm{c} = \left( \frac{1-T_\mathrm{a}}{1+T_\mathrm{a}} \right)^2 
\label{eq:Rc48}
\\
T_\mathrm{c} = 1- R_\mathrm{c} 
\;\; &\Rightarrow \;\;
T_\mathrm{c} = 
\frac{4 \sqrt{R_\mathrm{1a}}}{(1+\sqrt{R_\mathrm{1a}})^2} =
\frac{4 T_\mathrm{a}}{(1+T_\mathrm{a})^2} 
\label{eq:Tc48}
\end{align}
Note, how equation (\ref{eq:Tc48}) resembles equation (\ref{eq:Rc8}), and hence how the \textit{transmissivity} of the central mirror in the alternative design is calculated similarly to the \textit{reflectivity} of the central mirror in the original design.\\

\noindent By inserting conditions (\ref{eq:ta48}) and  (\ref{eq:rc48}) into Eq.~(\ref{eq:rcpaB}) and substituting $r_\text{1a} = \sqrt{R_\text{1a}}$, we obtain the following expression:
\begin{footnotesize}
\begin{equation}
    R_\text{CPA}^\text{(B)} =
    \frac{8 R_\text{1a} \cos^4(\phi)}{1 - 4 R_\text{1a}^{1/2} + 9 R_\text{1a} - 4 R_\text{1a}^{3/2} + R_\text{1a}^2 + \left( 1 - 4 R_\text{1a}^{1/2} + 2 R_\text{1a} - 4 R_\text{1a}^{3/2} + R_\text{1a}^2\right) \cos(2 \phi) + R_\text{1a}\cos(4 \phi) } \label{eq:RCPABopt}
\end{equation}
\end{footnotesize}

\noindent Note the similarity in the structure between expressions (\ref{eq:RCPABopt}) and (\ref{eq:RCPAAopt}). Expanding Eq.~(\ref{eq:RCPABopt}) around the resonance points $\phi_m = \frac{2m+1}{2} \pi$ reveals the anticipated quartic behavior of $R_\text{CPA}^\text{(B)}$:
\begin{equation}
    R_\mathrm{CPA}^\text{(B)}(\phi) = \frac{4 R_\text{1a}}{(\sqrt{R_\text{1a}}-1)^4} \left(\phi - \frac{2m+1}{2} \pi \right)^4 + \mathcal{O}^6\left(\phi - \frac{2m+1}{2} \pi \right)^6
\end{equation}

\section{S7. Comparison of Design A and Design B}

\begin{figure}[!ht]
\includegraphics[width=12cm]{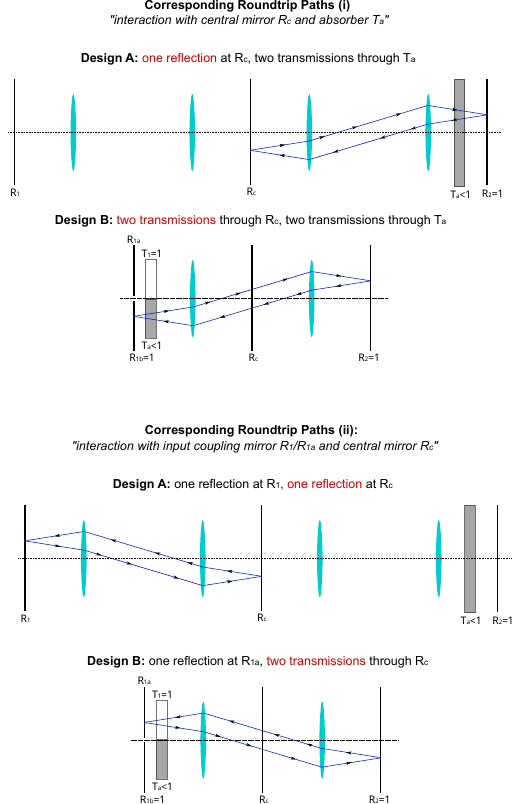}
\caption{\label{fig:comparison_1} Corresponding round-trip-paths (i) and (ii). In these paths, one single reflection at the central mirror in Design~A corresponds to two transmissions through the central mirror in Design~B.} 
\end{figure}

\noindent To approach the functionality of Design~B intuitively, it is instructive to compare its round-trip-paths with those in Design~A, and to account for each lossy reflection or transmission. This method allows us to exclude reflections at perfect mirrors from our consideration. In Design~A, light can take self-imaging round-trips in three ways: (i)~In the right sub-cavity, between the central mirror $R_{\mathrm{c}}$ and the right mirror $R_2$. Each round-trip involves one reflection at $R_{\mathrm{c}}$ and two passes through the absorber $T_\text{a}$. The equivalent in Design~B (where there is also an interaction with the central mirror and the absorber) is the circular path between bottom-left (mirror $R_{1\mathrm{b}}$) and top-right (mirror $R_2$), where light passes $T_\text{a}$ twice (as before), but also passes the central mirror twice. (ii)~Design~A also allows for round-trips in the left sub-cavity between $R_1$ and $R_\text{c}$. Here, each round-trip includes one reflection at $R_1$ and one reflection at $R_\text{c}$. The corresponding path in Design~B (where there is also an interaction with the input-coupling mirror and the central mirror) is the circular path between top-left (mirror $R_{1\mathrm{a}}$) and bottom-right (mirror $R_2$), featuring one reflection at $R_{1\text{a}}$ (as before), and again two passes through the central mirror per round-trip. Both round-trip-paths paths (i) and (ii) have in common that one single reflection at the central mirror in Design~A corresponds to two transmission through the central mirror in Design~B (see Fig.~\ref{fig:comparison_1})

\begin{figure}[!ht]
\includegraphics[width=13.5cm]{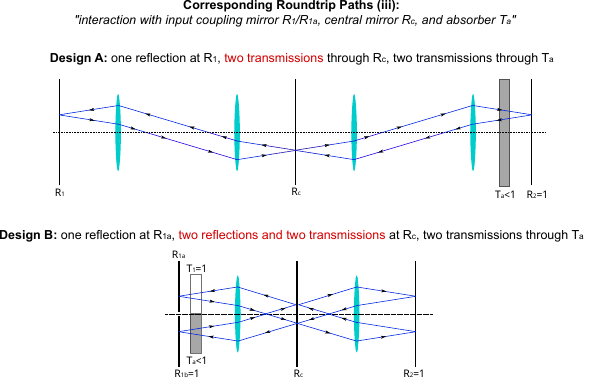}
\caption{\label{fig:comparison_2} Corresponding round-trip-paths (iii). In these corresponding paths, the light-field in both designs interacts with the input-coupling mirror, the central mirror, and the absorber. But now \textit{two transmissions} through the central mirror in Design~A correspond to \textit{two reflections and two transmissions} at the central mirror in Design~B.} 
\end{figure}

\noindent (iii) In Design~A, light also propagates between the farthest mirrors $R_1$ and $R_2$, getting reflected once at $R_1$, passing twice through the central mirror and twice through $T_\text{a}$. Design~B replicates this with a path following $R_{1\text{a}}$ -- $R_{\text{c}}$~-- $R_2$ (bottom) -- $R_\text{c}$ -- $R_2$ (top) -- $R_\text{c}$ -- $T_\text{a}$~-- $R_{1\text{b}}$~-- $T_\text{a}$ -- $R_{\text{c}}$ -- $R_{1\text{a}}$ (see Fig.~\ref{fig:comparison_2}). In this path (where the light also interacts with the input-coupling mirror, the central mirror, and the absorber) each round-trip involves one single reflection at  $R_{1\text{a}}$, two passes through $T_\text{a}$, two transmissions through the central mirror (in the bottom part), and two reflections at the central mirror (at the top part). In cases (i) and (ii), a \textit{single} reflection at the central mirror in Design~A is replaced by two transmissions through the central mirror in Design~B. This suggests that setting the transmissivity $T_\text{c}$ of the central mirror in Design~B equal to the square-root of the reflectivity $R_\text{c}$ in Design~A would establish an equivalence between the two designs. However, case (iii) complicates the matter: Here, \textit{two} transmissions in Design~A correspond to \textit{two transmissions and two reflections} in Design~B. At first glance, this appears to render Designs~A and B inequivalent. However, it is crucial to recognize that such comparisons between Design~A and Design~B offer, at best, a qualitative analogy. A comprehensive comparison requires the sound mathematical analysis from section S6, considering the infinite sum of all possible round trips and the effects of coupling between the respective sub-cavities. Nonetheless, this qualitative comparison indicates that the condition for the \textit{transmissivity} $T_\text{c}$ of the central mirror in Design~B may resemble the condition for the \textit{reflectivity} $R_\text{c}$ of the central mirror in Design~A, which is perfectly in accordance with the  result of the detailed mathematical analysis in section S6. 

\noindent Finally, it is important to note that positioning the absorber next to mirror $R_\mathrm{1b}$ in Design~B is not the only viable option. Effective EP configurations can also be achieved by placing the absorber in the right sub-cavity, adjacent to the central coupling mirror  $R_\mathrm{c}$ or next to to mirror $R_2$, either above or below the optical axis. 

\section{S8. Compensation of refraction aberrations}

\noindent This section delineates the methodology for compensating the refraction aberrations induced by the weak absorber placed between the rightmost lens and the rightmost mirror $R_2$ in Design~A, as illustrated  in Fig.~1(d) of the main paper, or adjacent to Mirror $R_{1\text{b}}$ in Design~B, as illustrated  in Fig.~4(a) of the main paper. We postulate that this absorber possesses a specific thickness, denoted as $d$, and  its refractive index's real component is represented by $n_r$. In the ensuing section, we detail the refraction compensation based on Design A. However, the findings are equally applicable to Design B.\\

\begin{figure}[ht]
\includegraphics[width=12cm]{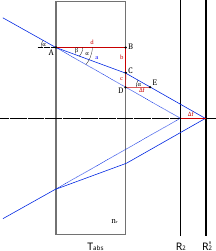}
\caption{\label{fig:coverslip_correction} The absorber of thickness $d$ with the real part of its refractive index being $n_r$ causes a shift $\Delta f$ in the focal point of the rightmost lens depicted in Fig.~1(d) of the main paper} 
\end{figure}

\noindent When the absorber is introduced without additional adjustments, it effectively shifts the focal point of the rightmost lens by a distance $\Delta f$ towards the right, a phenomenon graphically represented in  Fig.~\ref{fig:coverslip_correction}. Neglecting this effect leads to the spatial degeneracy of the MAD-EP-CPA being compromized. \\

\noindent The preliminary step in formulating a compensation strategy for this diffractive aberration is to determine the precise value of $\Delta f$. This is a simple ray-optics textbook problem, and can easily be achieved with some trigonometry. In the first step, we observe that the angle $\beta$ is functionally dependent on the incidence angle $\alpha$, as deduced from the fundamental principles of the law of refraction: 
\begin{align}
    n_\text{r} = \frac{\sin(\alpha)}{\sin(\beta)} \;\; \Longrightarrow \;\;
    \beta = \text{arcsin}\left( \frac{\sin(\alpha)}{n_\text{r}}\right) \label{eq:step1}
\end{align}
\noindent Subsequently, by employing the geometrical properties of triangle $ABC$, the distance $b$ can be mathematically represented as a function of the parameters $d$ and $\beta$:
\begin{align}
    \tan(\beta) = \frac{b}{d}
    \;\; \Longrightarrow \;\;
    b = d \, \tan \left( \beta \right) \label{eq:step2a}
\end{align}
Inserting (\ref{eq:step1}) into (\ref{eq:step2a}) leads to:
\begin{align}
     b = d \, \tan \left(  \text{arcsin}\left( \frac{\sin(\alpha)}{n_\text{r}}\right)  \right) 
     \;\; \Longrightarrow \;\;
     b = d \, \frac{\sin(\alpha)}{\sqrt{n_{\text{r}}^2-\sin^2(\alpha)}}
     \label{eq:step2b}
\end{align}
In the third step, the framework of triangle $ABD$ is utilized to formulate the sum of distances $b$ and $c$ as a function of $\alpha$ and $d$. Subsequently, this equation is transformed to isolate and express the variable $c$:
\begin{align}
    \tan(\alpha) = \frac{b+c}{d}
    \;\; \Longrightarrow \;\;
    b+c = d \, \tan(\alpha) 
    \;\; \Longrightarrow \;\;
    c = d \, \tan(\alpha) - b    \label{eq:step4}
\end{align}

\noindent Inserting (\ref{eq:step2b}) into this expression, allows us to express $c$ as a function of $\alpha$:
\begin{align}
    c = d \, \left[ \tan(\alpha) - \frac{\sin(\alpha)}{\sqrt{n_\text{r}^2-\sin^2(\alpha)}}\right]  \label{eq:step5}
\end{align}
In the final step, through the application of the properties of triangle $CDE$, a relationship is established between the variables $\Delta f$, $c$, and $\alpha$:
\begin{align}
    \tan(\alpha)=\frac{c}{\Delta f}
    \;\; \Longrightarrow \;\;
    \Delta f = \frac{1}{\tan(\alpha)} \, c \label{eq:step6}
\end{align}
Inserting (\ref{eq:step5}) into this expression finally leads to:
\begin{align}
    \Delta f =  d\,\left[ 1 - \frac{\cos(\alpha)}{\sqrt{n_\text{r}^2-\sin^2(\alpha)}}\right]
\end{align}
For small incident angles $\alpha$ this can be approximated as 
\begin{align}
    \Delta f \approx d \left( 1-  \frac{1}{n_\text{r}}\right) \label{eq:Deltaf}
\end{align}
The issue with this result lies in the practical limitations associated with adjusting the position of the rightmost mirror in the MAD-EP-CPA setup, as illustrated in Fig.~1(d) of the main text. Altering the mirror's position by a distance $\Delta f$, while maintaining the focal length $f$ of the adjacent lens constant, results in different optical lengths of the left and the right sub-cavity. While this modification corrects spatial degeneracy, it also changes the right sub-cavity's length, introducing several challenges: (i)~For effective operation as an MAD-EP-CPA, a specific longitudinal resonance mode of the left sub-cavity must be chosen. It is then essential to ensure that $\Delta f$ is an integer multiple of the corresponding wave-length. (ii) Unlike a symmetric design where both cavities are of equal optical length and the EP-CPA exhibits EP behavior across all longitudinal resonance frequencies, this asymmetric design is limited to a single wavelength. Calculations reveal that (keeping all other parameters equal) even a $\Delta f$  as small as $1 \text{mm}$  can result in a minimum absorption of about $1\%$ at the longitudinal mode adjacent to the resonant wavelength, with minimum absorption deteriorating rapidly for all further neighboring modes.  \\

\noindent Therefore, it is more advantageous to establish a new focal length, referred to as $f'$,  for the rightmost lens, such that the total optical path length between the two focal points of this lens remains unchanged. Let us define $L$ as the optical path length between the two focal points of the rightmost lens prior to the modification, and $L'$ as the optical path length between these points post-modification. The optical path length before the modification can be succinctly expressed as follows:
\begin{align}
    L = 2f  \label{eq:Lopt}
\end{align}
The optical path length $L'$ subsequent to the modification  can be represented as a function of the physical length $L_\text{phys}$. The physical length needs to be adjusted to account for the different optical path length in the absorber (compared to air), which can be expressed as $d(n_\text{r}-1)$. Consequently, we can express the optical length after the modification as:
\begin{align}
    L' = L^{'}_\text{phys} + d(n_\text{r}-1)  \label{eq:Lsopt1a}
\end{align}
The physical length after the modification can be quantified as 
\begin{align}
    L^{'}_\text{phys} =2f' + \Delta f  \label{eq:Lsphys}
\end{align}
Consequently, by inserting Eq.~(\ref{eq:Lsphys}) into Eq.~(\ref{eq:Lsopt1a}), we can express the optical length after the modification as:
\begin{align}
    L' = 2f' + \Delta f + d(n_\text{r}-1)  \label{eq:Lsopt1}
\end{align}
Inserting Eq. (\ref{eq:Deltaf}) into Eq. (\ref{eq:Lsopt1}) leads to:
\begin{align}
    L' = 2f' + d\left( n_\text{r} - \frac{1}{n_\text{r}}\right)  \label{eq:Lsopt2}
\end{align}
We want the optical path length to remain unchanged after the modification. Therefore we stipulate:
\begin{align}
    L' = L  \label{eq:L=L}
\end{align}
Inserting (\ref{eq:Lsopt2}) and (\ref{eq:Lopt}) into Eq. (\ref{eq:L=L}) finally leads to:
\begin{align}
    f' = f - \frac{d}{2}\left(n_\text{r} - \frac{1}{n_\text{r}} \right) \label{eq:fs}
\end{align}
Therefore, adjusting the focal length $f$ of the rightmost lens in the MAD-EP-CPA configuration, as illustrated in Fig.~1(d) of the main text, to a new value $f'$ in accordance with Eq.~(\ref{eq:fs}), effectively compensates for refraction aberrations, reinstates spatial degeneracy, and preserves the exceptional point.\\

\noindent Note that Eq.~(\ref{eq:fs}) is also applicable for preventing refraction aberrations in the alternative design (``Design B''), as depicted in Fig.~4 of the main paper.
\pagebreak


\par\noindent\rule{\textwidth}{0.5pt}

~~\\

\noindent [1] A. E. Siegman, Lasers \textit{(Revised)} (UNIVERSITY SCIENCE BOOKS, 1986).

\noindent [2] J. Frei, X.-D. Cai, and S. Muller, Multiport s-parameter and t-parameter conversion

with symmetry extension, IEEE Transactions on Microwave Theory and Techniques \textbf{56}, 

2493 (2008).

\noindent [3] Y. Slobodkin, G.Weinberg, H. H\"orner, K. Pichler, S. Rotter, and O. Katz, Massively

 degenerate coherent perfect absorber for arbitrary wavefronts, Science \textbf{377}, 995 (2022). 

\noindent [4] H. van de Stadt and J. M. Muller, Multimirror fabry–perot interferometers, Journal of 

the
Optical Society of America A \textbf{2}, 1363 (1985).

\noindent [5] F. A. Graybill, \textit{Matrices with applications in statistics} (Belmont, Calif. : Wadsworth

International Group, 1983).

\noindent [6] T.-T. Lu and S.-H. Shiou, Inverses of 2 × 2 block matrices, Computers \& Mathematics 

with Applications 43, 119 (2002).